\documentclass[11pt]{article}
\usepackage[latin1]{inputenc}
\usepackage{amsmath}
\usepackage{amsfonts}
\usepackage{amssymb,amsthm,multirow}
\usepackage{bm}
\usepackage[lined,boxed,linesnumbered]{algorithm2e}
	\DontPrintSemicolon
\usepackage{graphicx,subfigure}
\usepackage{url}
\usepackage{natbib}
\usepackage{array}
\usepackage{epstopdf} 
\usepackage{ulem}

\renewcommand \baselinestretch{1.5}

\def\bmu{\mbox{\boldmath{$\mu$}}}
\newcommand{\bOmega}{\mbox{\boldmath{${\Omega}$}}}
\newcommand{\bGamma}{\mbox{\boldmath{${\Gamma}$}}}
\newcommand{\bbeta}{\mbox{\boldmath{$\beta$}}}

\newcommand{\bgamma}{\mbox{\boldmath{${\gamma}$}}}
\newcommand{\balpha}{\mbox{\boldmath{${\alpha}$}}}

\newcommand{\bLambda}{\mbox{\boldmath{${\Lambda}$}}}
\def\bSigma{\mbox{\boldmath{$\Sigma$}}}

\def\smt{{\mbox{\tiny T}}}

\def\bX{\mathbf X}
\def\bx{{\bf x}}
\def\by{{\bf y}}
\def\bz{{\bf z}}

\def\bSb{\mathbf{S}_{b}}
\def\bSw{\mathbf{S}_{w}}

\def\bu{\mathbf u}
\def\bv{{\bf v}}
\def\bbe{\mathbf e}
\def\bbf{{\bf f}}

\def\bA{\mathbf A}
\def\bB{\mathbf B}
\def\bU{\mathbf U}

\def\bX{\mathbf X}
\def\bx{\mathbf x}
\def\bI{\mathbf I}

\def\bS{\mathbf S}

\def\bV{{\bf V}}

\def\bD{\mathbf D}

\def\hbmu{\mbox{\boldmath{$\hat{\bmu}$}}}

\usepackage{color}
\usepackage{comment}
\usepackage{ulem}
\usepackage{soul}

\newtheorem{thm}{Theorem}

\newtheorem{prop}{Proposition}

\def\rr{\textcolor[rgb]{1,0,0}}

\def\rb{\textcolor[rgb]{0,0,1}}

\newtheorem{rmk}{Remark}

\newcommand{\blind}{0}

\graphicspath{{images/}}
\begin{document}

\def\spacingset#1{\renewcommand{\baselinestretch}%
	{#1}\small\normalsize} \spacingset{1.3}


\if0\blind
{
	\title{\bf Sparse Linear Discriminant Analysis for Multi-view Structured Data }
	\author{Sandra E. Safo$^{1*}$, Eun Jeong Min$^2$, and Lillian Haine$^1$ \\
\textit{ssafo@umn.edu, mineunj@pennmedicine.upenn.edu, haine108@umn.edu}\\
		$^1$Division of Biostatistics,	University of Minnesota, Minneapolis, MN\\
		$^2$Department of Biostastics, Epidemiology and Informatics\\
		University of Pennsylvania, Philadelphia, PA
	}
	\maketitle
} \fi

\begin{abstract}
\noindent
Classification methods that leverage the strengths of data from multiple sources (multi-view data) simultaneously have enormous potential to yield more powerful findings than two step methods: association followed by classification. We propose two methods, sparse integrative discriminant analysis (SIDA) and SIDA with incorporation of network information (SIDANet), for joint association and classification studies. The methods consider the overall association between multi-view data, and the separation within each view in choosing discriminant vectors that are associated and optimally separate subjects into different classes. SIDANet is among the first methods to incorporate prior structural information in joint association and classification studies. It uses the normalized Laplacian of a graph to smooth coefficients of predictor variables, thus encouraging selection of predictors that are connected and behave similarly. We demonstrate the effectiveness of our methods on a set of synthetic and real datasets. Our findings underscore the benefit of joint association and classification methods if the goal is to correlate multi-view data and to perform classification.\\

\noindent$^*$\textit{corresponding author}\\

\end{abstract} 
\clearpage
\section{Introduction}\label{sec:intro} 
The problem of assessing associations among $d\ge 2$ data from multiple sources (also called multi-view data) measured on the same subject and assigning that subject into one of $K\ge 2$ classes based on multiple predictor variables from these views of data is an important  problem in modern biomedical research. With advancements in technologies, multiple diverse but related high-throughput data such as gene expression, metabolomics and proteomics data,  are often times measured on the same subject. A common research goal is to effectively synthesize information from these sources of data in order to identify factors (e.g., genetic and environmental [e.g., metabolites ]) that potentially separate subjects into different  groups. Many applications exist that consider this important problem \citep{langley2013integrated,lloyd2019multi}. Popular approach in the literature for integrative analysis and/or classification studies can broadly be grouped into three categories: association, classification, or joint association and classification methods. The literature on the first two is numerous, but the literature on the latter is rather limited. We focus on developing integrative analysis and classification methods to identify multi-view variables that are highly associated and optimally separate subjects into different groups. 

\subsection{Motivating Application}
Our work is motivated by a scientific  need to identify ``nontraditional" biomarkers (e.g., genes, metabolites) predictive of atherosclerosis cardiovascular diseases (ASCVD) beyond established risk factors (such as age and gender). Cardiovascular diseases (including ASCVD) continue to be the leading cause of death in the U.S and have become the costliest chronic disease \citep{CVD:2016}. The medical costs for CVD in 2016 was about $ \$1$ billion/day. It is projected that nearly half of the U.S. population will have some form of cardiovascular disease by 2035 and will cost the economy about $\$2$ billion/day in medical costs \citep{CVD:2016}.  Established environmental risk factors for CVD (e.g., age, gender, hypertension) account for only half of all cases of CVD \citep{Bartels:2012}. Finding other novel risk factors of CVD unexplained by traditional risk factors is important and may help prevent cardiovascular diseases. Trans-omics integrative analysis can leverage the strengths of omics to further our understanding of the molecular architecture of CVD. Since the metabolome is considered the end product of all genomic, epigenetic, and environmental activities \citep{Griffin2006,KRUMSIEK2016198}, linking metabolite levels in human plasma with gene expression data can identify multi-omics biomarkers predictive of ASCVD, and potentially serve as targets for interventions. 

We integrate gene expression, metabolomics, and/or clinical data from the Emory University and Georgia Tech Predictive Health Institute (PHI) study. The PHI study, which began in 2005, is a longitudinal study of healthy employees of Emory University and Georgia Tech aimed at collecting health factors that could be used to recognize, maintain, and optimize health rather than to treat  disease.   To  advance  this  goal,  we  seek  to leverage the strengths of multi-omics data in classification methods to  identify potential biomarkers beyond established risk factors that can distinguish between subjects at high-vs low- risk for developing ASCVD in 10 years.

\subsection{Existing Methods}
As mentioned earlier, the literature for integrative analysis and/or classification studies can be broadly grouped into three categories: association, classification, or joint association and classification methods. Association-based methods correlate multiple views of data to identify important variables as a first step. This is followed by independent classification analyses that use the identified variables. The techniques for correlating these multi-view data can be univariate or multivariate. The univariate approach considers  variables from one view as the response (e.g., each protein variable as response) and variables from the other views as predictors (e.g., one genetic variant) with a focus on one variable (e.g., one protein and one genetic variant) at a time. This approach is limiting since larger sample size is usually needed to identify associated variables, which is costly. Additionally, univariate methods assume variables within each view are independent and take no consideration of the dependency structure among variables. The multivariate techniques, on the other hand, assume variables within and between the views are dependent and use dimension reduction methods to simultaneously correlate multiple variables within and across multiple views \citep{Hotelling:1936,WT:2009,SAFOBIOM2018,EJ2019}. 
The association-based methods, either univariate or multivariate, are still largely disconnected from the classification procedure and oblivious of the effects class separation have on the overall dependency structure.

The classification-based methods either stack the views and perform classification on the stacked data, or individually use each view in classification algorithms and the results pooled. Several classification methods, including Fishers linear discriminant analysis (LDA)\citep{Fisher:1936} and its variants \citep{HBT:1995,BL:2004,GHT:2007,WT:2011,CHWE:2011,CL:2011,Shao:2011,SA:2015,Irina:2015,SAFOSADM2019}, support vector machines \citep{Cortes1995}, and random forest \citep{Breiman2001} may be used. These techniques take no consideration of the dependency structure between the views, and may be computationally expensive when the dimension of each view is large. 

Finally, the joint association- and classification-based methods \citep{WT:2009,kan2015multi,CCAReg2016,li2018integrative, zhang2018joint} link the problem of assessing associations between multiple views to the problem of classifying subjects into one of two or more groups within each view. The goal is then to 
identify linear combinations of the variables in each view that are correlated with each other  and have high discriminatory power. Limited literature exists for joint association- and classification- based methods.  \cite{WT:2009} introduced a supervised approach to  canonical correlation analysis (CCA), where the canonical correlation vectors were used to  predict a binary response in the CCA optimization problem. \cite{CCAReg2016} considered a regression formulation of CCA and proposed a joint method for obtaining the canonical correlation vectors and predicting an outcome using the canonical correlation vectors.  Their method is only applicable to binary classification problems. In addition, although the method is developed for multi-view data, the software they provide can only be used when there are two views of data. Recently, \cite{zhang2018joint} proposed a joint association and classification method that combines  linear discriminant and canonical correlation analysis using the regression formulation of these methods. Their method is useful for multi-class classification problems. The method we propose in this paper  falls into this category. 

\subsection{Overview of the proposed methods}
Our proposal is related to existing joint association- and classification-based methods but our contributions are multi-fold. First, we also consider joint association and classification problems, but our formulation of the problem is different from the regression approach largely considered by existing methods; this provides a  different insight into the same  problem. We directly solve the optimization problem of maximizing association and separation of classes using Lagrangian methods, resulting in systems of eigenvalue-vector problems that is easily solved. 
More importantly, our methods rely on summarized data (i.e., covariances) making them applicable if the individual view cannot be shared due to privacy concerns.  Secondly, while existing association and classification methods concentrate on sparsity (i.e., exclude nuisance predictors), which is mainly data-driven, our SIDANet method is both data- and knowledge-driven. SIDANet uses the normalized Laplacian of a graph to smooth the rows of the discriminant vectors for each view, thus encouraging predictors that are connected and behave similarly to be selected or neglected together. The benefits of excluding nuisance predictors have been widely acknowledged in the statistical literature and these include better interpretability, improved classification or prediction estimates, and computational efficiency \citep{Tibshirani:1994,Dantzig:2007}. Incorporating prior knowledge about variable-variable interactions has the potential of identifying functionally meaningful variables (or network of variables) within each view for   improved  classification performance. This approach has been successful in several applications including regression \citep{LL:2008,PXS:2010}, classification \citep{SAFOSADM2019}, and association studies \citep{Chenetal:2013,SAFOBIOM2018}. Thirdly, our formulation makes it easy to include other covariates without enforcing sparsity on the  coefficients corresponding to the covariates. This is rarely done in integrative analysis and classification methods. Including other available covariates may inform the choice of variables to be excluded, which in turn may result in better classification estimates.  Fourth, our formulation of the problem can be solved easily with any off-the-shelf convex optimization software. We develop computationally efficient algorithms that take advantage of parallelism. Table 1 highlights the unique features of our proposed methods compared to existing works. 

The rest of the paper is organized as follows. In Section 2, we briefly discuss the motivation of our proposed methods. In Section 3, we present the proposed methods for two views of data. In Section 4, we introduce the sparse versions of the proposed methods. In Section 5, we extend the proposed methods to more than two views of data.  In Section 6, we present the algorithm for implementing the proposed methods. In Section 7, we present how the discriminant vectors from the proposed methods could be used for classification. In Section  8, we conduct simulation studies to assess the performance of our methods in comparison with other methods in the literature. In Section 9, we apply our proposed methods to a real data. We conclude with some discussion remarks in Section 10. 
\begin{table} \label{tab:uniquefeatures}
	\begin{small}
		\begin{tabular}{lllllll}
			\hline
			\hline
			Property/&Classification-&  Association-&	JACA&CCA-& \textbf{SIDA}& \textbf{SIDANet}\\
			Method&Based&  Based&	~&Regression& ~& ~\\
			\hline
			\hline
			Association & &  \checkmark& \checkmark & \checkmark & \checkmark & \checkmark  \\
			\hline
			Classification& \checkmark& & \checkmark & \checkmark* & \checkmark & \checkmark  \\
			\hline
			Variable Selection& \checkmark&  \checkmark& \checkmark & \checkmark & \checkmark & \checkmark  \\
			\hline
			Smoothness&\checkmark & \checkmark & &  & &\checkmark  \\
			\hline
			Covariates& & &  &  & \checkmark &\checkmark \\
			\hline
			\hline
		\end{tabular}
		\caption{Unique features of SIDA and SIDANet compared to other methods. *CCA-regression is not applicable when there are more than two classes.}
	\end{small}
\end{table}

\section{Motivation}
Suppose there are two sets of high-dimensional data  $\bX^{1} =(\bx^1_{i},\cdots,\bx^1_{n})^{\smt} \in \Re^{n \times p}$ and  $\bX^{2} = (\bx^2_{i},\cdots,\bx^2_{n})^{\smt} \in \Re^{n \times q}$, $p,q > n$, all measured on the same set of subjects, $i=1,\ldots,n$. For subject $i$, let $y_{i}$ be the class $k$ ( $k=1,\ldots K$) membership. Given these data, we wish to predict the class membership $y_{j}$ of a  new subject $j$  using their high-dimensional information $\bz_{j}^{1} \in \Re^{p}$ and $\bz_{j}^{2} \in \Re^{q}$. Several supervised classification methods, including Fishers linear discriminant analysis (LDA)\citep{Fisher:1936}, support vector machines \citep{Cortes1995}, random forest \citep{Breiman2001} may be used to predict class membership when there is only one view of data, but not when there are two views of data.  On the other hand, unsupervised association methods, including canonical correlation analysis (CCA) \citep{Hotelling:1936} and co-inertia analysis \citep{COI1994} could be used to study association between the two views of data, but are not suitable when classification is the ultimate goal. We propose two methods for joint association and classification problems that bridge the gap between LDA and CCA. We use the LDA formulation in our problem. Although some of the aforementioned classifiers have demonstrated remarkable predictive performances, many of the predictions from these methods are not interpretable \citep{SU2018,Doshi2017}. In many biomedical research problems, just knowing a single metric, such as classification accuracy, is not enough; an emphasis is  also placed on specific features that lead to the classification estimates.  We briefly describe LDA and CCA for completeness sake.

\noindent\textit{\textbf{\large{Linear Discriminant Analysis}}}\\
For the description of LDA, we suppress the superscript in $\bX$. 
Let ${\bX_{k}}=(\bx_{1},\ldots,\bx_{n_{k}}), \bx \in \Re^{p}$ be the data matrix for class $k$, $k=1,\ldots,K$, and $n_{k}$  the number of samples in class $k$. Then, the mean vector for class $k$,  common covariance matrix for all classes, and the between-class covariance are respectively given by 
\[\hat{\bmu}_{k} = (1/n_{k})\sum_{i=1}^{n_{k}}\bx_{ik} ;~~ \bSw = \sum\limits_{k=1}^{K}\sum\limits_{i=1}^{n}(\bx_{i}-\hat{\bmu}_{k})(\bx_{i}-\hat{\bmu}_{k})^{\smt};  ~~~~\bSb = \sum\limits_{k=1}^{K}n_{k}(\hat{\bmu}_{k}-\hat{\bmu})(\hat{\bmu}_{k}-\hat{\bmu})^{\smt}. \nonumber\]
Here, $\hat{\bmu}$ is the combined class mean vector and is defined as $\hbmu=(1/n)\sum\limits_{k=1}^{K}n_{k}\hbmu_{k}$.
For a $K$ class prediction problem, LDA finds $K-1$ direction vectors, which are linear combinations of all available variables, such that projected data have maximal separation between the classes and minimal separation within the classes. Mathematically, the solution to the  optimization problem:
\begin{eqnarray} \label{eqn:fisherkint}
\max_{\bbeta_{k}} \bbeta_{k}^{\smt}\bSb \bbeta_{k} ~~~\mbox{subject to}~~ \bbeta_{k}^{\smt}\bSw\bbeta_{k} =1,
~~\bbeta_{l}^{\smt}\bSw\bbeta_{k} =0~~\forall l<k,~~k=1,2,\dots,K-1
\end{eqnarray}
yields the LDA directions that optimally separate the $K$ classes and these are the eigenvalue-eigenvector pairs $(\hat{\lambda}_{k},\hat{\bbeta}_{k})$, $\hat{\lambda}_{1}>\cdots> \hat{\lambda}_{k}$ of $\bSw^{-1}\bSb$ for $\bSw \succ 0$. The data are then projected onto the LDA directions to obtain the LDA scores $(\bX\hat{\bbeta}_{1},\ldots,\bX\hat{\bbeta}_{K-1})$. These scores could be visualized for separation patterns. 
\begin{figure}\label{fig:LDAgoal}
	\begin{tabular}{ll}
		\includegraphics[height = 2.2in]{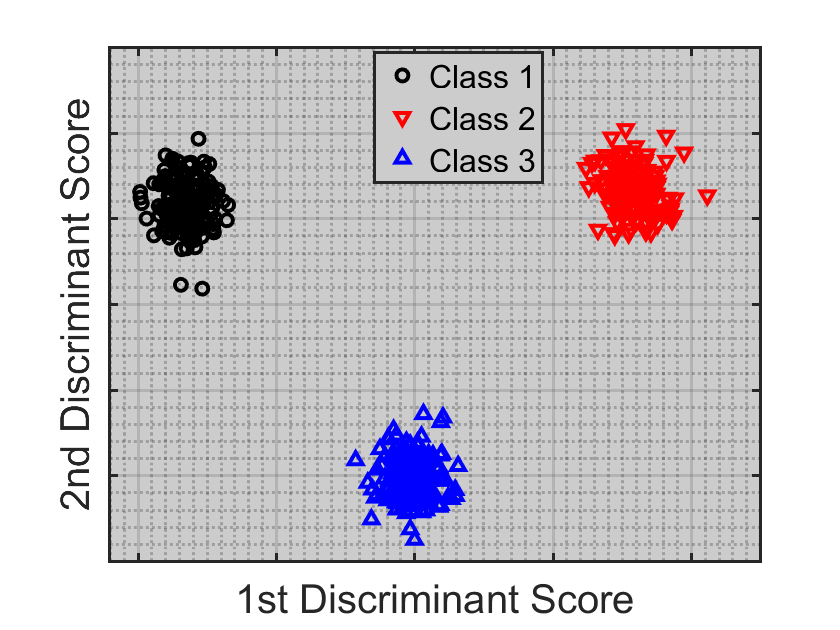} & 	
		\includegraphics[height = 2.2in]{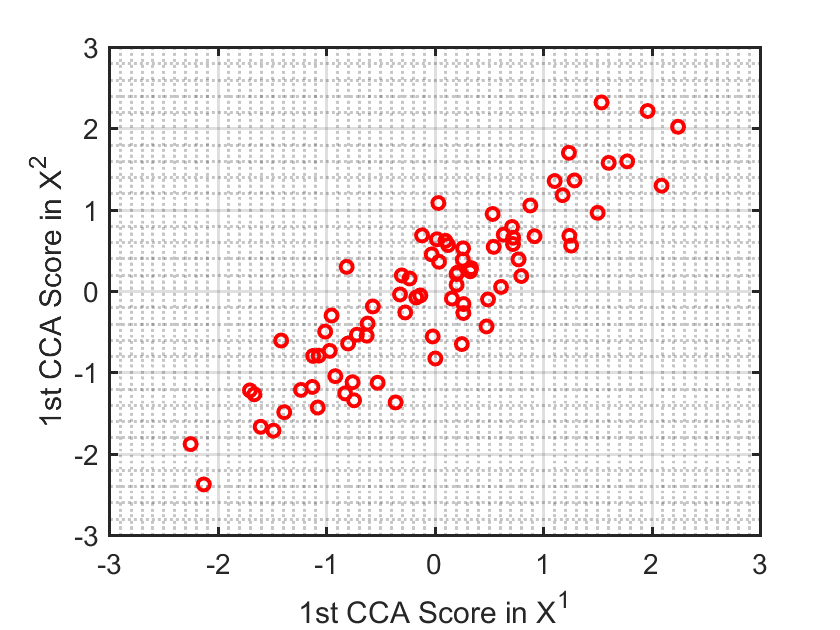} \\
	\end{tabular}
	\caption{Projection plot of a) LDA and b) CCA scores.}
\end{figure} 
Figure 1 a) is a projection plot of data onto the two LDA directions for a $K=3$ class problem.

\noindent\textit{\textbf{\large{Canonical Correlation Analysis}}}\\
Unlike the classical LDA which is useful for assessing separation between classes for either $\bX^{1}$ or $\bX^{2}$, canonical correlation analysis (CCA) may be used for assessing associations between $\bX^{1}$ and $\bX^{2}$. Without loss of generality, we assume  $\bX^{1}$ and $\bX^{2}$ have zero means for each variable.  
The goal of CCA \citep{Hotelling:1936} is to find linear combinations of the variables in $\bX^{1}$, say $\bX^{1} \balpha$ and in $\bX^{2}$, say $\bX^{2} \bbeta$, such that the correlation between these linear combinations is maximized.
If $\bS_{1}$ and $\bS_{2}$ are  sample covariances of $\bX^{1}$ and $\bX^{2}$ respectively, and $\bS_{12}$ is the $p \times q$ sample cross-covariance  between $\bX^{1}$ and $\bX^{2}$, then  mathematically, CCA finds   $\balpha$ and $\bbeta$ that solves the optimization problem:
\begin{eqnarray} \label{eqn:ccaopt}~~~~~~
\max_{\balpha,\bbeta} ~\balpha^{\smt}\bS_{12}\bbeta~~~\mbox{subject to}~~ \balpha^{\smt}\bS_{1}\balpha =1~~ \mbox{and~~}\bbeta^{\smt}\bS_{2}\bbeta =1.
\end{eqnarray}
The  solution to problem (\ref{eqn:ccaopt}) is given as $\hat{\balpha}=\bS_{1}^{-1/2}\bbe_{1}, \hat{\bbeta}=\bS_{2}^{-1/2}\bbf_{1}$, where $\bbe_{1}$ and $\bbf_{1}$ are the first left and right singular vectors of $\bS_{1}^{-1/2}\bS_{12}\bS_{2}^{-1/2}$. Once the first CCA directions have been obtained, the data are then projected to these directions to visualize the strength of association between the two data types. Figure 1 b) is a projection plot of the first CCA direction for $\bX^{1}$ and $\bX^{2}$.\\
\noindent\textit{\textbf{\large{Our proposed approach}}}\\
We propose a method that combines both LDA and CCA. 
Specifically, we  1) maximize (a) the sum of the between class separations of the two views of data, and (b) the squared correlations between the views and 2) allow for only important  variables or networks of variables to contribute to the overall  association and separation. 
In the next section, we describe our technique for obtaining integrative discriminant (IDA) directions for two views of data. In Section 4, we discuss our approach for achieving sparse integrative discriminant (SIDA) directions. In Section 5, we extend the proposed problem to more than two views of data. 

\section{Discriminant analysis for two views of data}
Consider a $K$-class classification problem with two sets of variables $\bX^{1} \in \Re^{n \times p}$ and $\bX^{2} \in \Re^{n \times q}$ and the class membership vector $\by$. Let $\bS_{12}$ be the covariance between $\bX^{1}$ and $\bX^{2}$.
Our goal is to find linear combinations of $\bX^{1}$ and $\bX^{2}$ that explain the overall association between these views while optimally separating  the K classes within each view. These optimal discriminant vectors could be used to effectively classify a new subject into one of the K classes using their available  data. We propose to solve the optimization problem below for $\tilde{\bA}=[\tilde{\balpha}_{1}, \ldots,\tilde{\balpha}_{k}]$ and $\tilde{\bB}=[\tilde{\bbeta}_{1}, \ldots,\hat{\bbeta}_{k}]$, $k=1,\ldots,K-1$:
\begin{eqnarray} \label{eqn:fisherkintMulti3}
\max_{\bA, \bB}  \rho\overbrace{\text{tr}(\bA^{{\smt}}\bS_{b}^{1} \bA+  \bB^{{\smt}}\bS_{b}^{2} \bB)}^{{\rr{\text{\large{separation}}}}}  &+& (1-\rho)\overbrace{ \text{tr}(\bA^{{\smt}}\bS_{12}\bB\bB^{{\smt}}\bS_{12}^{\smt}\ \bA)}^{{\rb{\text{\large{association}}}}}
\nonumber\\
~~~~\mbox{subject to}~~ \text{tr}(\bA^{{\smt}}\bS_{w}^{1} \bA)/(K-1)&=&1 ,~~  \text{tr}(\bB^{{\smt}}\bS_{w} \bB)/(K-1) =1.
\end{eqnarray}  
Here, tr($\cdot$) is the trace function, and $\rho$ is a parameter that controls the relative importance of the separation and association terms in the objective.
The first term in equation (\ref{eqn:fisherkintMulti3}) considers the separation between classes within each view and the second term considers the association between the two views of data through the squared correlation. Essentially, the goal here is to uncover some basis directions that influence both separation and association. Consider optimizing the  problem above using Lagrangian multipliers. One can show that the solution reduces to a set of generalized eigenvalue problems. Theorem \ref{thm:GEVPmain}
gives a formal representation of the solution to the optimization problem (\ref{eqn:fisherkintMulti3}).

\begin{thm}\label{thm:GEVPmain}
	Let $\bS_{w}^{1}, \bS_{w}^{2}$ and $\bS_{b}^{1}, \bS_{b}^{2}$ respectively be within-scatter and between-scatter covariances for $\bX^{1}$ and $\bX^2$. Let $\bS_{12}$ be the covariance between the two views of data. Assume $\bS_{w}^{1} \succ 0$,  $\bS_{w}^{2} \succ 0$.  
	Then $\bA=(\balpha_{1},\ldots,\balpha_{r})^{\smt} \in \Re^{p \times r}$, $\bB=(\bbeta_{1},\ldots,\bbeta_{r})^{\smt} \in \Re^{q \times r}, k=1,\ldots, r$ are eigenvectors corresponding respectively to eigenvalues $\bLambda_{1}=$diag$(\lambda_{1_{k}},\ldots,\lambda_{1_{r}})$ and $\bLambda_{2}=$diag$(\lambda_{2_{k}},\ldots,\lambda_{2_{r}})$, $\lambda_{1_{k}} > \cdots > \lambda_{1_{r}}>0$, $\lambda_{2_{k}} > \cdots > \lambda_{2_{r}}>0$ that iteratively solve the generalized eigenvalue (GEV) system:
	\begin{eqnarray}\label{eqn:GEViter}
	(\rho\bS_{b}^{1} + \rho\bS_{b}^{1^{\smt}} + (1-\rho) \bOmega^1 + (1-\rho)\bOmega^{1^{\smt}} )\bA&=&(\bS_{w}^{1}+\bS_{w}^{1^{\smt}}) \bLambda_{1}\bA \\
	(\rho\bS_{b}^{2} + \rho\bS_{b}^{2^{\smt}} + (1-\rho)\bOmega^2+ (1-\rho)\bOmega^{2^{\smt}} )\bB&=&(\bS_{w}^{2}+\bS_{w}^{2^{\smt}}) \bLambda_{2}\bB \
	\end{eqnarray} 
\end{thm}
\noindent where $\bOmega^1=\bS_{12}\bB\bB^{\smt}\bS_{12}^{\smt} $ and $\bOmega^2=\bS_{12}^{\smt}\bA\bA^{\smt}\bS_{12}$.
Equations (\ref{eqn:GEViter}) and (5) may be solved iteratively by fixing $\bB$ and solving an eigensystem for $\bA$, and then fixing $\bA$ and solving an eigensystem in (5) for $\bB$. The algorithm may be initialized using any arbitrary normalized nonzero vector. With $\bB$ fixed at $\bB^*$ in (4), the solution is the eigenvalue-eigenvector pair of $(\bS_{w}^{1}+\bS_{w}^{1^{\smt}})^{-1}(\rho\bS_{b}^{1} + \rho\bS_{b}^{1^{\smt}} + (1-\rho)\bOmega^1 + (1-\rho)\bOmega^{1^{\smt}} )$. Similarly, with $\bA$ fixed at $\bA^*$ in (5), the solution of (5) is the  eigenvalue-eigenvector pair of $(\bS_{w}^{2}+\bS_{w}^{2^{\smt}})^{-1}(\rho\bS_{b}^{2} + \rho\bS_{b}^{2^{\smt}} + (1-\rho)\bOmega^2+ (1-\rho)\bOmega^{2^{\smt}} )$. 

\begin{rmk}
	$\xi_1= {\bX^1}\balpha_1$  and $\eta_1= {\bX^2}\bbeta_1$  are two linear combinations with  variances $1$ having the maximum separation and squared correlation among joint separations and correlations between any two linear combinations ${\bX^1}\balpha$ and ${\bX^2}\bbeta$.
\end{rmk}
\begin{rmk}
	\textit{Rank determination}. In the classical LDA problem, the rank (maximum number of eigenvalues) is $K-1$, where $K$ is the number of classes. This coincides with  \text{rank}($\bS^1_{b}$) (or \text{rank}($\bS^2_{b}$) ). For a fixed $\bB^*$, $rank\left((\bS_{w}^{1}+\bS_{w}^{1^{\smt}})^{-1}(\bS_{b}^{1} + \bS_{b}^{1^{\smt}} + \bOmega^1 + \bOmega^{1^{\smt}} )\right)$
	\begin{eqnarray}
	&\le& K-1 + \min\left(\text{rank}(\bS_{w}^{1^{-1}}), \text{rank}(\bS_{12}),\text{rank}(\bB) \right) \nonumber.
	\end{eqnarray}
	This suggests that for the integrative LDA problem, there are more than $K-1$ eigenvalue-eigenvector pairs. In  practice, one could use a scree-plot to choose the rank. However, in our simulations and real data analyses, we find that the first $K-1$ eigenvalues dominate the rest of the eigenvalues. Thus, we set the maximum number of eigenvalues to be $K-1$, similar to the classical LDA.  
\end{rmk}
\begin{rmk}
	Note that if the two views of data are weakly correlated so that $\bS_{12}$ is negligible, then the $k$-th eigenvalues $\lambda^1_k$ and $\lambda^2_k$ from integrative LDA will coincide with the eigenvalues obtained from separate applications of original LDA on  $\bX^{1}$ or $\bX^{2}$. Hence, there will not be any advantage to an integrative LDA. 
\end{rmk}

We rewrite the optimization problem  (\ref{eqn:fisherkintMulti3}) and the generalized eigensystems (4) and (5) in equivalent forms to facilitate computations. We omit it's proof for brevity sake since it follows easily from (\ref{eqn:fisherkintMulti3}). Let $\mathcal{M}^1=\bS_w^{1^{-1/2}}\bS_{b}^{1}\bS_w^{1^{-1/2}}$, $\mathcal{M}^2=\bS_w^{2^{-1/2}}\bS_{b}^{2}\bS_w^{2^{-1/2}}$. Also, let $\mathcal{N}_{12}=\bS_w^{1^{-1/2}}\bS_{12}\bS_w^{2^{-1/2}}$ and $\mathcal{N}_{21}=\bS_w^{2^{-1/2}}\bS_{12}^{\smt}\ \bS_w^{1^{-1/2}}$.
{\prop\label{prop:GEVeq}
	\begin{small}
		The maximizer (\ref{eqn:fisherkintMulti3}) is equivalent to $(\widetilde{\bA}, \widetilde{\bB})= (\bS^{1^{-1/2}}_{w}\widetilde{\bGamma^{1}}, \bS^{2^{-1/2}}_{w}\widetilde{\bGamma^{2}})$ where
		\begin{eqnarray*} \label{eqn:intequiv}
			(\widetilde{\bGamma^{1}},\widetilde{\bGamma^{1}})=\max_{\bGamma^1, \bGamma^2}  \rho\text{tr}(\bGamma^{{1\smt}}\mathcal{M}^1 \bGamma^1+  \bGamma^{{2\smt}}\mathcal{M}^2 \bGamma^2) 
			&+& (1-\rho)\text{tr}(\bGamma^{{1\smt}}\mathcal{N}_{12}\bGamma^2\bGamma^{{2\smt}}\mathcal{N}_{21}\bGamma^1)\nonumber\\
			~~~~\mbox{subject to}~~ \text{tr}(\bGamma^{{1\smt}}\bGamma^1)/(K-1)&=&1 ,~~  \text{tr}(\bGamma^{{2\smt}}\bGamma^2)/(K-1) =1
		\end{eqnarray*}  
		
		Furthermore, this yields the equivalent eigensystem problems of (4) and (5)
		\begin{eqnarray}\label{eqn:GEVitereq}
		(\rho\mathcal{M}^1 + \rho\mathcal{M}^{1^{\smt}} + (1-\rho)\bar{\mathcal{N}}_{12} + (1-\rho)\bar{\mathcal{N}}_{12}^{{\smt}} )\bGamma^1&=& \bLambda_{1}\bGamma^1 \nonumber\\
		(\rho\mathcal{M}^{2} + \rho\mathcal{M}^{2^{\smt}} + (1-\rho)\bar{\mathcal{N}}_{21}+ (1-\rho)\bar{\mathcal{N}}_{21}^{{\smt}} )\bGamma^2&=& \bLambda_2\bGamma^2\
		\end{eqnarray} 
		where $\bar{\mathcal{N}}_{12}=\mathcal{N}_{12}\bGamma^2\bGamma^{2\smt}\mathcal{N}_{21}$ and $\bar{\mathcal{N}}_{21}=\mathcal{N}_{21}\bGamma^1\bGamma^{1\smt}\mathcal{N}_{12}$.
	\end{small}
}

\begin{rmk}
	In high-dimensional examples where $p > n$, we make $\bSw^1$ and $\bSw^2$ positive definite by adding a small multiple of the identity. We could estimate $\bSw^1$ and $\bSw^2$ using techniques proposed in \cite{CLL:2011} and \cite{BickelLevina2008} but that would add a layer of complexity.  To reduce computations, we use techniques described in \cite{HastieTrevor:2004} to  avoid inverting the $p \times p$ (or $q \times q$) matrices $\bS^{1^{1/2}}_{w}$ and $\bS^{2^{1/2}}_{w}$; instead, we invert a $n \times n$ matrix, and $n \ll p$ (or $q$). 
\end{rmk}

\section{Sparse LDA for two views of data }
The linear discriminant vectors that solve the joint association and classification problem (\ref{eqn:intequiv}) are especially useful in the low-dimensional settings where $n > p$ since it yields direction vectors that are easily interpretable. In the high-dimensional setting where $n\ll p$, $\bGamma^1$ and $\bGamma^2$ are weight matrices of all available variables in $\bX^{1}$ and $\bX^{2}$. These coefficients are not usually zero (i.e., not sparse) making interpreting the discriminant functions challenging. 
We propose to  make $\bGamma_1$ and $\bGamma_2$ sparse by imposing convex penalties  subject to modified eigensystem constraints. Our approach follows ideas in \cite{SafoBiomSELP}, which is in turn motivated by the Dantzig selector \citep{Dantzig:2007}.  We impose penalties that depend on whether or not prior knowledge in the form of functional relationships are available or not. 

In what follows, for a vector $\textbf{v} \in \mathbb{R}^p$ we define $\|\textbf{v}\|_\infty = \max_{i=1,\cdots,p}|v_i|$, $\|\textbf{v}\|_1 = \sum_{i=1}^p |v_i|$, and $\|\textbf{v}\|_2 = \sqrt{\sum_{i=1}^p v_i^2}$. For a matrix $\textbf{M} \in \mathbb{R}^{n \times p}$ we define $\textbf{m}_i$ to be its $i$th row, $\textbf{m}_j$ to be its $j$th column, and define the maximum absolute row sum $\|\textbf{M} \|_\infty =\max_{1\le i\le n}\sum_{j=1}^p|m_{ij}| $.

\subsection{Sparse Integrative Discriminant Analysis (SIDA)}
Let $\bGamma^1=(\bgamma_{1}^1,\ldots,\bgamma_{p}^1)^{1\smt} \in \Re^{p \times K-1}$ and $\bGamma^2=(\bgamma_{1}^2,\ldots,\bgamma_{q}^2)^{\smt} \in \Re^{q \times K-1}$ denoate the collection of basis vectors that solve the eigen systems (\ref{eqn:intequiv}). To achieve sparsity, we define the following block $l_1/l_2$  penalty functions that consider the length of row elements in  $\bGamma^1$ and $\bGamma^2$ and shrinks the row vectors of irrelevant variables to zero:
\begin{equation}\label{eqn:pen}
\mathcal{P}(\bGamma^d)= \sum_{i=1}^{p ~\text{or}~q}\|\bgamma_{i}^d\|_{2},~~~d=1,2 
\end{equation}
We note that variables with null effects are encouraged to have zero coefficients simultaneously in all basis directions. This is because the block $l_1/l_2$ penalty applies the $l_2$-norm $\|\bgamma_{i}^d\|_2$ within each variable, and the $l_1$-norm across variables,   and thus shrinks the row length to zero. This results in coordinate-independent variable selection, making it appealing for screening irrelevant variables. With penalty (\ref{eqn:pen}), we obtain sparse solutions $\widehat{\bGamma}^1$ and $\widehat{\bGamma}^2$ by iteratively solving the following convex optimization problems for fixed $\bGamma^1$ or $\bGamma^2$:
\begin{small}
	\begin{eqnarray}\label{eqn:joint}
	\min_{\bGamma^1} \sum_{i=1}^{p}\|\bgamma_{i}^1\|_{2} \quad &\mbox{s.t}& \quad \|(\rho\mathcal{M}^1 + \rho\mathcal{M}^{1^{\smt}} + (1-\rho)\bar{\mathcal{N}}_{12} + (1-\rho)\bar{\mathcal{N}}_{12}^{{\smt}} )\widetilde{\bGamma}^1- \widetilde{\bLambda}_{1}\bGamma^1 \|_\infty \leq\tau_{1}\nonumber\\
	\min_{\bGamma^2} \sum_{i=1}^{q}\|\bgamma_{i}^2\|_{2} \quad &\mbox{s.t}& \quad \|(\rho\mathcal{M}^2 + \rho\mathcal{M}^{2^{\smt}} + (1-\rho)\bar{\mathcal{N}}_{21} + (1-\rho)\bar{\mathcal{N}}_{21}^{{\smt}} )\widetilde{\bGamma}^2 - \widetilde{\bLambda}_2\bGamma^2\|_\infty \leq\tau_{2}. 
	\end{eqnarray}
\end{small}
Equation (\ref{eqn:joint}) essentially constrains the first and second eigensystems (\ref{eqn:intequiv})  to be within $\tau_1$ and $\tau_2$ respectively. It can be easily shown that naively  constraining the  eigensystems result in trivial solutions. Hence, we substitute $\bGamma^1$ and $\bGamma^2$ in the left hand side (LHS) of the eigensystem problems in (\ref{eqn:intequiv}) respectively with $\widetilde{\bGamma^1}$ and  $\widetilde{\bGamma^2}$, the nonsparse solutions that solve equation (\ref{eqn:intequiv}). We choose to substitute the LHS instead of the right hand side (RHS) equation in (\ref{eqn:intequiv}) because we are able to recover the nonsparse solutions when $\tau_1 =0$ and $\tau_2=0$. Additionally, we obtain numerically stable solutions with the LHS substitution. Here, $(\widetilde{\bLambda}_{1}, \widetilde{\bLambda}_{2})$ are the eigenvalues corresponding to $\widetilde{\bGamma}^1$ and $\widetilde{\bGamma}^2$. Also,  $(\tau_{1}, \tau_{2})$ are tuning parameters controlling the level of sparsity; their selection will be discussed in Section 6. $\widehat{\bGamma^1}$ may be obtained from (\ref{eqn:joint}) by fixing $\bGamma^2$ (definition of $\bar{\mathcal{N}}_{12}$ involves $\bGamma^2$). Similarly, $\widehat{\bGamma^2}$ may be obtained by fixing $\bGamma^1$. The solutions $(\widehat{\bGamma^1}, \widehat{\bGamma^2})$ may not necessarily be orthogonal, as such we use Gram-Schmidt orthogonalization on $(\widehat{\bGamma^1}, \widehat{\bGamma^2})$.

\begin{rmk}
	Inclusion of covariates: Our optimization problems in (\ref{eqn:joint}) make it easy to include other covariates to potentially guide the selection of relevant variables  likely to improve classification accuracy. Assume that  $\tau_2$ is set to zero (no penalty on the corresponding coefficients). Then $\widetilde{\bGamma}^2$ solves the second optimization problem. But the basis discriminant directions $\widehat{\bGamma}^1$ for the first view of data depend on the second view ($\bX^{2}$) through the correlation matrix $\bS_{12}$. Thus, to account for the influence of covariates in the optimal basis discriminant directions, one could always include the available covariates (as a separate view) and set corresponding tuning parameter to zero. This forces data from the covariates to be used in assessing associations and separations without necessarily shrinking their effects to zero. For binary (e.g, age) or categorical covariates (assumes no ordering), we suggest the use of indicator variables \citep{gifi1990nonlinear}. All variables are standardized to have  zero and variance one, so that any dominant effect of a variable on the correlation matrices is not due to the unit of measurement.  
\end{rmk}

\subsection{Sparse Integrative Discriminant Analysis (SIDA) for structured data (SIDANet)}
We introduce SIDANet for structured or network data. SIDANet utilizes prior knowledge about variable-variable interactions in the estimation of the sparse integrative discriminant vectors.  For instance, in biomedical research, information about variable connectivity may be obtained from networks such as protein-protein networks, biochemical networks, transcription regulation networks, and metabolic-metabolic networks. Many databases exist for obtaining such information about variable-variable relationships. One such database for protein-protein interactions is the human protein reference database (HPRD) \citep{hprd2003}. 
We capture the variable-variable connectivity within each view in our sparse discriminant vectors via the normalized Laplacian \citep{chung1997spectral} obtained from the underlying graph. 
Let $\mathcal{G}^d =(V^d,E^d,W^d)$, $d=1,2$ be a network given by a weighted undirected graph. $V^d$ is the set of vertices corresponding to the $p^d$ variables (or nodes) for the $d$-th view of data. Let $E^d=\{u \sim v\}$ if there is an edge from variable $u$ to $v$ in the $d$th view of data. Let $r_v$ denote the degree of the vertex $v$ (i.e., the number of variables connected to node $v$) within each view. $W^d$ is the weight of an edge for the $d$-th view satisfying $w(u,v) =w(v,u) \ge 0$. Note that if $\{u,v\} \not \in E(\mathcal{G})$, then $w(u,v)=0$.
The normalized Laplacian of $\mathcal{G}^d$ for the $d$-th view is 
\begin{eqnarray} \label{eqn:laplacian}
\mathcal{L}_n(u,v)  &=& \begin{cases}
1-w(u,v)/r_v & \mbox{if ~$u$~$=$~$v$~and ~$r_v \ne 0$} \\ 
-\frac{w(u,v)}{\sqrt{r_u r_v}} & \mbox{if $u \ne v$ and  ~variables~$u$~ and ~$v$~ are~adjacent} \\ 
0 &~ \mbox{otherwise}.\
\end{cases} 
\end{eqnarray}
The matrix $\mathcal{L}_n(u,v)$ is usually sparse (has many zeros) and so can be stored with sparse functions in any major software programs such as R or Matlab. For smoothness while incorporating prior information, we impose the following penalty: 
\begin{equation}\label{eqn:sidanet}
\mathcal{P}(\bGamma^d)=\eta\sum_{i=1}^{p^d}\|\bgamma^{\mathcal{L}_n}_{i}\|_{2} + (1-\eta) \sum_{i=1}^{p^d}\|\bgamma_{i}\|_{2}. 
\end{equation}
$\bgamma^{\mathcal{L}_{n}}_{i}$ is the $i$-th row of the matrix product $\mathcal{L}_n\bGamma^d$. Note that  $\mathcal{L}_n(u,v)$ is different for each view. The first term in equation (\ref{eqn:sidanet}) acts as a smoothing operator for the weight matrices $\bGamma^d$ so that variables that are connected within the $d$-th view are encouraged to have a similar effect, and so would be selected or neglected together. The second term in equation (\ref{eqn:sidanet}) enforces sparsity of variables within the network; this is ideal for eliminating variables or nodes that contribute less to the overall association and discrimination relative to other nodes within the network. $\eta$ balances these two terms. 

\begin{rmk}
	One could use the Laplacian (not normalized) defined as:
	\begin{eqnarray} \label{eqn:rlaplacian}
	\mathcal{L}(u,v)  &=& \begin{cases}
	r_v-w(u,v) & \mbox{if ~$u$~$=$~$v$} \\ 
	-w(u,v) & \mbox{if $u \ne v$ and  ~variables~$u$~ and ~$v$~ are~adjacent} \\ 
	0 &~ \mbox{otherwise}\
	\end{cases} 
	\end{eqnarray}
	instead of the normalized Laplacian defined in equation (\ref{eqn:laplacian}). However, the Laplacian in equation (\ref{eqn:rlaplacian}) encourages variables in the network to have the same effect size (coefficients). This is true since $w(u,v)$ is the same for variables that are connected. We believe  variables that are connected will often have different coefficients or effect sizes that capture their contributions to overall dependency structure and class separation. As such, we use the normalized Laplacian, which normalizes the connected variables by their degrees, thus encouraging different effect sizes. 
\end{rmk}

\section{Extension to multiple views of data}
We extend the proposed method to more than two views of data. Let $\bX^d=[\bX^d_{1}, \bX^d_{2},\ldots,\bX^{d}_{K}]$, $\bX^{d}\in \Re^{n\times p_d}, \bX^{d}_{k} \in \Re^{n_k \times p_d}, k=1,\ldots,K,~ d=1,2,\ldots,D$ be a concatenation of the $K$ classes in the $d$-th view. Let $\bSb^{d}$ and $\bSw^d$ be the between-class and within-class covariances for the $d$-th view. Let  $\bS_{dj}, j<d$ be the cross-covariance between the $d$-th and $j$-th views. Define $\mathcal{M}^d=\bS_w^{d^{-1/2}}\bS^d_b\bS_w^{d^{-1/2}}$ and 
$\mathcal{N}_{dj}=\bS_w^{d^{-1/2}}\bS_{dj}\bS_w^{j^{-1/2}}$. We solve the optimization problem for multiple views of data: 
\begin{small}
	\begin{eqnarray*} \label{eqn:intequiv2}
		\max_{\bGamma^d}  \rho\sum_{d=1}^D\text{tr}(\bGamma^{{d\smt}}\mathcal{M}^d \bGamma^d) 
		+ \frac{2(1-\rho)}{D(D-1)}\sum_{d=1,d\ne j }^{D}\text{tr}(\bGamma^{{d\smt}}\mathcal{N}_{dj}\bGamma^j\bGamma^{{j\smt}}\mathcal{N}_{jd}\bGamma^d)
		~\mbox{s.t}~\text{tr}(\bGamma^{{d\smt}}\bGamma^d)=K-1.
	\end{eqnarray*}  
\end{small}
\noindent As before, $\rho$ controls the influence of separation or association in the optimization problem.  The second term essentially sums all of these pairwise squared correlations and weight them by $\frac{D(D-1)}{2}$ so that the sum of the squared correlations  is one.  As in proposition \ref{prop:GEVeq}, the nonsparse basis discriminant directions for the $d$-th view, $\widetilde{\bGamma}^d$, are given by the eigenvectors corresponding to the eigenvalues that iteratively solve the following eigensystems:

\begin{eqnarray}\label{eqn:GEVitermultiple}
\left({c}_1\mathcal{M}^{1} + {c_2} \sum_{j\ne 1}^{D}\mathcal{N}_{1j}\bGamma^{j}\bGamma^{j^{\smt}}\mathcal{N}_{j1}\right)\bGamma^1&=&\bLambda_{1}\bGamma^1, \nonumber\\
\vdots\nonumber\\
\left({c_1}\mathcal{M}^{D} + {c_2}\sum_{j=1 }^{D-1}\mathcal{N}_{Dj}\bGamma^{j}\bGamma^{j^{\smt}}\mathcal{N}_{jD}\right)\bGamma^D&=&\bLambda_{D}\bGamma^D,
\end{eqnarray} 
where we set ${c_1}=\rho$ and ${c_2}=\frac{2(1-\rho)}{D(D-1)}$.
For sparsity or smoothness we solve the following optimization problems:
\begin{eqnarray}\label{eqn:joint2}
\min_{\bGamma^1} \mathcal{P}(\bGamma^1) \quad &\mbox{s.t}& \quad \|(c_1\mathcal{M}^1 + c_1\mathcal{M}^{1^{\smt}} + c_2\bar{\mathcal{N}}_{1j} + c_2\bar{\mathcal{N}}_{1j}^{{\smt}} )\widetilde{\bGamma}^1- \widetilde{\bLambda}_{1}\bGamma^1 \|_\infty \leq\tau_{1}\nonumber\\
\vdots\nonumber\\
\min_{\bGamma^D} \mathcal{P}(\bGamma^D) \quad &\mbox{s.t}& \quad \|(c_1\mathcal{M}^D + c_1\mathcal{M}^{D^{\smt}} + c_2\bar{\mathcal{N}}_{Dj} + c_2\bar{\mathcal{N}}_{Dj}^{{\smt}} )\widetilde{\bGamma}^D - \widetilde{\bLambda}_D\bGamma^D\|_\infty \leq\tau_{D} 
\end{eqnarray}
where $\bar{\mathcal{N}}_{dj}=\sum_{d,j }^{D}\mathcal{N}_{dj}\bGamma^{j}\bGamma^{j^{\smt}}\mathcal{N}_{jd}$, $d,j=1,\ldots D$, and $j\ne d$ sums all pairwise correlations of the $d$-th and the $j$-th views. The penalty term $\mathcal{P}(\bGamma^d)$ is either set respectively to equation (\ref{eqn:pen}) or (\ref{eqn:sidanet}) depending on whether sparsity or smoothness (with sparsity) is desired. 

\section{Initialization, tuning parameters, and algorithm}
The optimization problems in equations (\ref{eqn:joint}) and (\ref{eqn:joint2}) are biconvex. With $\bGamma^d$ fixed  at $\bGamma^{d^*}$, the problem of solving for $\widehat{\bGamma}^j$, $j\ne d$ is convex, and may be  solved easily with any-off-the shelf convex optimization software.  The technique of solving biconvex problems by fixing parameters and then solving the resulting convex problems is  popularly used in the statistical literature. At the first iteration, we fix $\bGamma^{d^*}$ as the classical LDA solution from applying LDA on $\bX^d$. We can initiate $\bGamma^{d^*}$ with  random orthonormal matrices, but we choose to initialize with regular LDA solutions because the algorithm converges faster. At subsequent solutions, we fix $\bGamma^{d^*}$ as the solution from previous iteration, and iterate until convergence. 
Algorithm \ref{alg::sida} gives an outline of our proposed methods.

The optimization problems depend on tuning parameters $\tau_d$, which need to be chosen. We fix $w=0.5$ to provide equal weight on separation and association. Without loss of generality, assume the $D$-th (last) view is the covariates, if available. We fix $\tau_D=0$ and select the optimal tuning parameters for the other views from a range of tuning parameters. Note that searching the tuning parameters hyperspace can be computationally intensive. For instance, if there are two views (excluding covariates) each having 10 grid points, then one needs to search a $10 \times 10$ grid space, representing 100 grid values to choose the optimal combination. For $d=1,..,D-1$, we need to search a large hyperparameter space [$(G_1 \times G_2\times\cdots\times G_{D-1})$ grid values assuming $G_d$ is the number of grid points for the $d$-th view]. This obviously is computationally taxing. To overcome this computational bottleneck, we follow ideas in \cite{bergstra2012random} and randomly select some grid points (from the entire grid space) to search for the optimal tuning parameters; we term this approach \textit{random search}. This technique has been shown to yield good results \citep{bergstra2012random} when compared to searching the entire space (\textit{grid search}). In fact, our own simulations with \textit{random search} produced satisfactory results (see Tables 2-5) when compared to \textit{grid search}. In our simulations and real data applications, for two views (excluding covariates), we set 8 grid points each, and randomly select $20\%$ of the grid values in the hyperparameter space to optimize. For $d>2$, we set the number of grid points to 5, and randomly select $15\%$ of the grid values in the hyperparameter space to optimize. A detailed comparison of \textit{random search}  and \textit{grid search} in terms of error rates, estimated correlations, variables selected, and computational time is found in Section 8 and the web supplemental material.  

We provide  upper and lower bounds for $\tau_d$. Let $d=1$.  Note that $\tau_1 > \| (c_1\mathcal{M}^1 + c_1\mathcal{M}^{1^{\smt}} + c_2\bar{\mathcal{N}}_{1j} + c_2\bar{\mathcal{N}}_{1j}^{{\smt}} ) \|_{\infty}$ results in trivial solution vectors, i.e., $\widehat{\bGamma}^1 = \bf{0}$. Hence, we set the upper bound for $\tau_1$ as $\tau_{1\max} =\| (c_1\mathcal{M}^1 + c_1\mathcal{M}^{1^{\smt}} + c_2\bar{\mathcal{N}}_{1j} + c_2\bar{\mathcal{N}}_{1j}^{{\smt}} ) \|_{\infty}$.
Similar results hold for the other views. Instead of using a lower bound of $0$, we use a lower bound dependent on the dimensions of each view (specifically $\tau_{d\min}= (\sqrt{\log{p^d}/n})\cdot\tau_{d\max})$ to encourage sparsity. We choose the optimal tuning parameters from the range of tuning parameters using $K$-fold cross validation ($K=5$ in our simulations and real data applications) to minimize average classification error. Our classification approach is found in Section \ref{sec:classify}.\\

\setlength{\intextsep}{2pt}
\begin{algorithm}[htb]
	\DontPrintSemicolon
	\SetKwComment{tcp}{$\triangleright\ $}{}
	\textbf{Input}: training data $(\bX^d,\mathbf{y})$; tuning parameters $\tau_d, d=1,\ldots,D$; edge matrix, $E^d$ and edge weight, $W^d$ (for SIDANet) \;
	\tcp*{{\small  $\tau_D=0$ if covariates ($D$-th view) available }}
	\textbf{Output}: estimated sparse discriminant vectors $\widehat{\bGamma}^d$. \;
	\textbf{Initialize}: $\bGamma^d$, $d=1,\ldots,D$ . \;
	\tcp*{{\small  Use random orthonormal matrices or solution from classical LDA }}
	\Repeat{ convergence}{
		\For{$d=1,\ldots,D$}{
			\begin{footnotesize}
				Fix $\widetilde{\bGamma}^d$ and $\widetilde{\bLambda}_d$.\;
				\tcp*{{\small  Use solutions from generalized eigenvalue systems  (equation \ref{eqn:GEVitermultiple}) }}\;
				Solve  \begin{eqnarray*}
					\min_{\bGamma^d} \mathcal{P}(\bGamma^d) \quad &\mbox{s.t}& \quad \|(c_1\mathcal{M}^d + c_1\mathcal{M}^{d^{\smt}} + c_2\bar{\mathcal{N}}_{dj} + c_2\bar{\mathcal{N}}_{dj}^{{\smt}} )\widetilde{\bGamma}^d - \widetilde{\bLambda}_d\bGamma^d\|_\infty \leq\tau_{d} 
				\end{eqnarray*} \\
				\tcp*{{\small  $\mathcal{P}(\bGamma^d)$ is defined in equation (\ref{eqn:pen}) for SIDA and (\ref{eqn:sidanet}) for SIDANet }}
			\end{footnotesize}
		}
	}
	\caption{Algorithm for obtaining sparse (and network-constrained) integrative discriminant vectors for multi-view data. }
	\label{alg::sida}
\end{algorithm}

\section{Using SIDA and SIDANet for classification}\label{sec:classify}
Once the SIDA or SIDANet discriminant functions have been obtained, one can make future class assignments by either 1) pooling the discriminant scores for each view $\bX^d, d=1\ldots D$, or 2) using individual discriminant scores from each view. The latter option, which we term separate class assignment, is appealing if for some reasons some of the views are not available for future observations. In such instances, future class assignments can be carried out using the discriminant functions for available views. In either the pooled or separate class assignments, we use nearest centroid for classification. 

\noindent The discriminant scores are defined to be $\bU^d=\bX^d\widehat{\bGamma}^d, d=1,...,D$, where $\widetilde{\bGamma}^d$ is a $p^d \times (K-1)$ matrix of basis vectors obtained from SIDA or SIDANet.  Let $\bz^d = (z_1^d,...,z_p^d)^\smt$  be the available measurement for a new (future) observation for the $d$-th view. Consider projecting these future observations onto the estimated discriminant vectors $\widehat{\bGamma}^d$ for the $d$-th view (i.e., $\bv^{d}=\bz^{d\smt}\widehat{\bGamma}^d$)  and concatenating the scores for all $d$ views; i.e $\bv=[\bz^{1\smt}\widehat{\bGamma}^1,\bz^{2\smt}\widehat{\bGamma}^2,\cdots,\bz^{D^\smt}\widehat{\bGamma}^D]^{\smt} \in \Re^{D(K-1)}$. For pooled class assignment, we assign $\bz=[\bz^1,\cdots,\bz^D]$ to class $k$ if the distance from $\bv$ to $\hat{\bu}_k$ is minimum, that is, 
\[\min_k \sum_{k=1}^{K}\|\bv - \hat{\bu}_k \|_2, ~~k=1,...,K\]
where $\hat{\bu}^{\smt}_k \in \Re^{D(K-1)}$ is the  pooled mean for class $k$ obtained from the pooled scores $\bU=[\bU^1,\cdots,\bU^D]\in \Re^{D(K-1)}$.
For separate class assignments, we assign $\bz^d$ to the population whose class mean is closest to $\bv^d$, i.e.,
\[\min_k \sum_{k=1}^{K}\|\bv^d - \hat{\bu}^d_k \|_2, ~~k=1,...,K, d=1,\cdots,D\]

\section{Simulations}\label{sec:simul}
We consider two main simulation examples to assess the performance of the proposed methods in identifying important variables and/or networks that optimally separate classes while maximizing association between multiple views of data. In the first example, we simulate a $D=2$, $K=2$ and $K=3$ class discrimination problem and assume there is no prior information available. In the second example, we simulate a $D=3$ and $K=3$ class problem and assume prior information is available in the form of networks. We focus on the situations where the true discriminant vectors are highly sparse in each view in order to test the ability of our methods in discovering signal variables when noise variables are also present. We consider different covariance structures, and partition the covariance matrix within each view into signal and noise; signals contain variables that are correlated and contribute to class separation within each view and overall association between views, while noise variables are uncorrelated and unimportant. In example two, we vary the structural information of the network so that all or some of the networks contribute to both separation and association. In each simulation example, 20 Monte Carlo datasets for each view are generated.

\subsection{Example 1: simulation settings when no prior information is available }\label{sec:simul1}
\normalsize{\textbf{Scenario One (Multi-class, equal covariance with class)}}: 
The first view of data  $\bX^{1}$ has $p$ variables and the second view  $\bX^{2}$ has $q$ variables, all drawn on the same samples with size $n=240$. Each view is a concatenation of data from three classes, i.e.,  $\bX^{d} = [\bX_1^d, \bX_2^d, \bX_3^d], d=1,2$. The combined data $\left(\bX^{1}_k, \bX^{2}_k\right)$ for each class are simulated from $N(\bmu_k, \bSigma)$, where $\bmu_k = (\bmu^1_k, \bmu^2_k)^{\smt} \in \Re^{p+q}, k=1,2,3$ is the combined mean vector for class $k$; $\bmu^1_k \in \Re^p, \bmu^2_k \in \Re^q$ are the mean vectors for $\bX^{1}_k$ and $\bX^{2}_k$ respectively. The true covariance matrix $\bSigma$ is partitioned as \begin{eqnarray}
\bSigma =\left(
\begin{array}{cc}
\bSigma^{1} & \bSigma^{12} \nonumber\\
\bSigma^{21} & \bSigma^{2} \nonumber\
\end{array} \right), \bSigma^1 =\left(
\begin{array}{cc}
\tilde{\bSigma}^{1} & \textbf{0} \nonumber\\
\textbf{0} & \bI_{p-20} \nonumber\
\end{array} \right), \bSigma^2 =\left(
\begin{array}{cc}
\tilde{\bSigma}^{2} & \textbf{0} \nonumber\\
\textbf{0} & \bI_{q-20} \nonumber\
\end{array} \right)
\end{eqnarray}

\noindent where $\bSigma^{1}$,  $\bSigma^{2}$ are respectively the covariance of  $\bX^{1}$ and $\bX^{2}$, and $\bSigma^{12}$ is the cross covariance between the two views. $\tilde{\bSigma}^{1}$ and $\tilde{\bSigma}^{2}$ are each block diagonal with 2 blocks of size 10, between-block correlation 0, and  each block is a compound symmetric matrix with correlation 0.7. We generate $\bSigma^{12}$ as follows.  Let $\mathbf{V}^{1} = [\mathbf{V}^{1}_{1},~ \mathbf{0}_{(p-20) \times 2}]^{\smt} \in \Re^{p \times 2 }$ where  the entries of $V^{1}_{1} \in \Re^{20 \times 2}$ are \textit{i.i.d} samples from U(0.5,1). We similarly define $\mathbf{V}^2$ for the second view,  and we  normalize  such that $\mathbf{V}^{1^{T}}\bSigma^{1}\mathbf{V}^{1} = \mathbf{I}$ and $\mathbf{V}^{2^{\smt}}\bSigma^2 \mathbf{V}^{2} = \mathbf{I}$. We then set  $\bSigma^{12} = \bSigma^{1}\bV^1\bD\bV^{2^{\smt}} \bSigma^{2}$, $\bD= \text{diag}(\rho_1, \rho_2)$. We vary $\rho_1$ and $\rho_2$ to measure the strength of the association between $\bX^{1}$ and $\bX^{2}$. For separation between the classes, we take $\bmu_k$ to be the columns of  $[\bSigma\bA, \textbf{0}_{p+q}]$, and $\bA=[\bA^1, \bA^2]^{\smt} \in \Re^{(p +q) \times 2}$. Here, the first column of $\bA^{1} \in \Re^{p \times 2}$ is set to $(c\textbf{1}_{10}, \textbf{0}_{p-10})$; 
\begin{figure}\label{fig:S1Allscenarios}
	\begin{tabular}{l}
		\includegraphics[height = 5.0in]{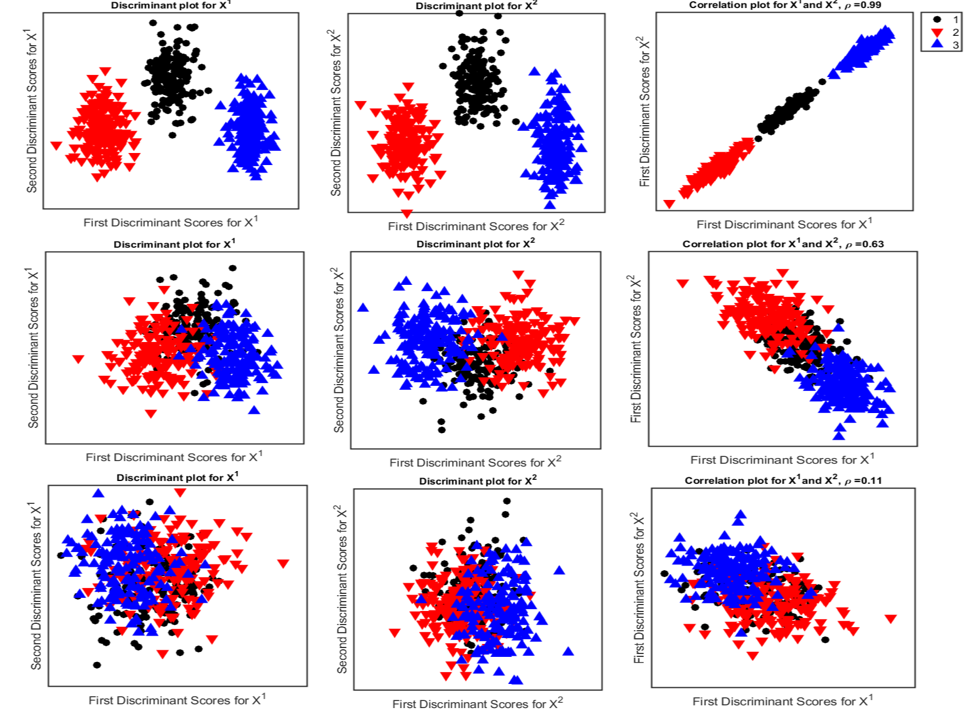}
	\end{tabular}
	\caption{Projection of random data simulated from Scenario One onto true integrative discriminant direction vectors. Top panel: good separation of classes, and strong association between views. Middle pane: moderate separation and moderate association. Bottom panel: weak separation and weak association. }
\end{figure} 
the second column is set to $( \textbf{0}_{10},-c\textbf{1}_{10}, \textbf{0}_{p-20})$. We set $\bA^{2}$ similarly. We vary $c$ to assess discrimination between the classes, and we consider three combinations of $(\rho_1, \rho_2, c)$  to assess both discrimination and strength of association. For each combination, we consider equal class size $n_k=80$, and dimensions $(p/q= 2000/2000)$. The true integrative discriminant vectors are the generalized eigenvectors that solve Theorem \ref{thm:GEVPmain}. Figure 2 is a visual representation of random data projected onto the true integrative discriminant vectors for different combinations of $c$, $p_1$ and $p_2$. In the top panel, $(\rho_1=0.9, \rho_2=0.7, c=0.5)$. In the middle panel, $(\rho_1=0.4, \rho_2=0.2, c=0.2)$. In the bottom panel, $(\rho_1=0.15, \rho_2=0.05, c=0.12)$.

\noindent\normalsize{\textbf{Scenario Two (Multi-class, unequal covariance within class)}}: 
In Scenario One we considered an example where the LDA assumption holds, i.e., the within-class covariance is the same for each class. In this setting, we relax this assumption. The covariance matrices for the three classes within $\bX^{1}$ and $\bX^{2}$ are each given as follows: for class 1, the covariance matrix has the same form as in Model 1; for class 2, the covariance matrix has entries $\sigma_{ij} = 0.6^{|i-j|}$; for class 3, the covariance matrix is the  identity matrix, $\textbf{I}_{(p ~\text{or} ~q)}$. \\

\noindent\normalsize{\textbf{Scenario Three (Binary class, equal covariance within class)}}: 
We consider a $D=2$ high-dimensional and $K=2$ class problem.  The covariance matrices for each class follow Scenario One. The mean matrices follow Scenario One but with this exception: $\bA^1 \in \Re^p$ is set to $(c\mathbf{1}_{20},\mathbf{0}_{p-20})$. $\bA^2$ is defined similarly. As before, we vary $c$ to assess separation between the two classes.  

\subsubsection{Competing Methods}
We compare SIDA with classification- and/or association-based methods. For the classification-based method, we consider MGSDA \citep{Irina:2015} and either apply MGSDA on the stacked data [MGSDA (Stack)], or apply MGSDA on separate datasets [MGSDA (Ens)]. To perform classification for MGSDA (Ens), we pool the discriminant vectors from the separate MGSDA applications, and apply the pooled classification algorithm discussed in Section 7. For association-based methods, we consider the sparse CCA (sCCA) method \citep{SafoBiomSELP}. We perform sCCA using the Matlab code the authors provide.  Similar to MGSDA (Ens), we perform classification for sCCA by pooling the canonical variates from CCA and applying the pooled classification algorithm discussed in Section 7.
We also compare SIDA to JACA \citep{zhang2018joint}, a method for joint association and classification studies. We use the  R package provided by the authors, and set the number of cross-validation folds as 5. We do not compare our method to the supervised sparse CCA \citep{WT:2009} and CCA regression \citep{CCAReg2016} methods because we have a three-class problem; these methods are only applicable to binary outcomes. 

\subsubsection{Evaluation Criteria}\label{sec:evalcriteria}
We evaluate the methods using the following criteria. (1) test  misclassification rate; (2) selectivity, and (3) estimated correlation. We consider three measures to capture the methods ability to select true signals while eliminating false positives: true positive rate (TPR), false positive rate (FPR), and $F_1$ score defined as follows: $TPR = \frac{TP}{TP +FN} $, $FPR = \frac{FP}{FP + TN}$, $F_1$ score$=  \frac{2 TP } {2TP + FP + FN}$, where TP, FP, TN, FN are defined respectively as true positives, false positives, true negatives, and false negatives. We estimate the overall correlation, $\hat{\rho}$, by summing estimated pairwise correlations obtained from the RV coefficient \citep{RVCoefficient}. The RV-coefficient for two centred matrices  $\mathcal{X} \in \Re^{n \times k}$ and $\mathcal{Y} \in \Re^{n \times k}$ is defined as 
$RV(\mathcal{X},\mathcal{Y}) = \frac{tr(\Sigma_{\mathcal{X}\mathcal{Y}}\Sigma_{\mathcal{Y}\mathcal{X}})}{\sqrt{tr(\Sigma^2_{\mathcal{X}\mathcal{X}})tr(\Sigma^2_{\mathcal{Y}\mathcal{Y}})}}$. The RV coefficient generalizes the squared Pearson correlation coefficient to multivariate data sets. We obtain the estimated correlation as
$\hat{\rho} = \frac{2}{D(D-1)} \sum_{d=1, d\ne j}^{D}RV({\bX^d_{test}\widehat\bGamma}^d, \bX^j_{test}\widehat{\bGamma}^j)$, $\hat{\rho} \in [0,1]$

\subsubsection{Results}
Tables 2 -4 show the averages of the evaluation measures from 20 repetitions, for scenarios one to three. We first compare SIDA with \textit{random search} [SIDA({RS})] to SIDA with \textit{grid search} [SIDA({RS})]. We note that across all evaluation measures, SIDA ({RS}) tends to be better or similar to SIDA (GS). In terms of computational time SIDA ({RS}) is faster than SIDA ({GS}) [refer to the web supplemental material]. This suggests that we can choose optimal tuning parameters at a lower computational cost by randomly selecting grid points from  the entire tuning parameter space and searching over those grid values, and  still achieve similar or even better performance compared to searching over the entire grid space.  We next compare SIDA with an association-based method, sCCA. In Scenario One, across all settings, we observe that SIDA (especially SIDA (RS)) tends to have a lower error rate, a comparable TPR, a lower FPR, and a higher $F_1$ score. The estimated correlation $\hat{\rho}$ is higher for sCCA in the settings where the correlation between the two views is moderate or weak, and the classes have more overlap.  Compared to a classification-based method, MGSDA (either Stack or Ens), SIDA has a lower error rate, higher estimated correlations (except in setting 3), higher TPR, and higher $F_1$ scores.  Similar results hold for Scenario Two (where we relax the assumption of equal covariances in each class) and Scenario Three (where we have a binary class problem). 
When compared to JACA, a joint association- and classification- based method, for Scenarios One and Three, SIDA has lower error rates in setting 1, and comparable error rates in settings 2 and 3. In terms of selectivity, SIDA has comparable TPR  in setting 1, lower TPR in setting 2, higher TPR in setting 3, lower or comparable FPR, comparable estimated correlations, and higher F$_1$ scores in settings 1 and 2. The performance for SIDA is slightly sub optimal in Scenario Two when compared to JACA.  

These simulation results suggest that joint integrative-and classification-based methods, SIDA and JACA, tend to outperform association- or classification-based methods. In addition, the proposed method, SIDA, tends to be better than JACA in the scenarios where the views are moderately or strongly correlated, and the separation between the classes is not weak.

\begin{table}[httb!]
	\begin{footnotesize}
		\begin{tabular}{lrrrrrrrr}
			\hline
			\hline
			Method&Error (\%)&  $\hat{\rho}$&	TPR-1&TPR-2 & FPR-1 &FPR-2  & F-1& F-2	\\
			\hline
			Setting 1 \\
			$(\rho_1=0.9, \rho_2=0.7, c=0.5)$\\
			SIDA (RS)	&0.04&  	0.99  	&100.00	&100.00	&0.00	&0.00	&100.00	&100.00\\
			SIDA (GS)	&0.05&0.99		&100.00	&100.00	&0.00	&0.00	&100.00	&100.00 \\
			sCCA	   &0.05&	0.99	    &100.00	&100.00	&1.04	&1.32	&69.89	&69.19\\
			JACA	   &0.11&	1.00	    &100.00	&100.00	&3.42	&3.86	&42.37	&38.07\\
			MGSDA (Stack)   &0.19 & 	0.84	&7.50	&8.50	&0.00	&0.00	&16.82	&16.20\\
			MGSDA (Ens)	   &0.33	&0.95	&14.25	    &13.50	&0.00	&0.05	&24.65	&22.46\\
			\hline
			Setting 2\\
			$(\rho_1=0.4, \rho_2=0.2, c=0.2)$\\
			SIDA (RS)&	11.32&	0.58&	100.00&	100.00&	1.17&	1.90&	86.56&	80.51\\
			SIDA (GS)&	11.42&0.58	&	100.00&	99.75&	2.28&	1.57&	68.82&	81.85\\
			sCCA&	16.20&	0.65&	100.00&	100.00&	2.44&	1.14&	66.70&	70.81\\
			JACA&	11.32&	0.58&	100.00&	100.00&	2.23&	1.94&	75.92&	76.38\\
			MGSDA (Stack)&	12.52&	0.55&	34.25&	32.50&	0.04&	0.06&	48.22&	46.29\\
			MGSDA (Ens)&    17.05&	0.61&	39.00&	37.00&	0.04&	0.07&	53.34&	50.09\\
			\hline
			Setting 3\\
			$(\rho_1=0.15, \rho_2=0.05, c=0.12)$\\
			SIDA (RS)	&31.03	&0.14	&98.50	&97.00	&5.07	&2.93	&41.43	&58.05	\\
			SIDA (GS)	&29.61	&0.26	&99.00	&99.75	&2.48	&2.85	&53.88	&56.07		\\
			sCCA	&34.80	&0.20	&92.75	&93.75	&1.10	&1.47	&74.66	&77.45\\
			JACA	&29.84	&0.19	&97.25	&97.00	&0.74	&0.85	&81.51	&82.53		\\
			MGSDA (Stack)	&31.55	&0.15	&28.00	&27.00	&0.07	&0.05	&41.53	&40.25	\\
			MGSDA (Ens)	&35.31	&0.16	&30.75	&28.50	&0.17	&0.01	&41.92	&43.09		\\
			\hline
			\hline
		\end{tabular}
		\caption{Scenario One:  RS; randomly select tuning parameters space to search. GS; search entire tuning parameters space. MGSDA (Ens) applies sparse LDA method on separate views and perform classification on the pooled discriminant vectors. MGSDA (Stack) applies sparse LDA on stacked views. TPR-1; true positive rate for $\bX^1$. Similar for TPR-2. FPR; false positive rate for $\bX^2$. Similar for FPR-2; F-1 is F-measure for $\bX^1$. Similar for F-2. $\rho_1$ and $\rho_2$ controls the strength of association between $\bX^1$ and $\bX^2$. $c$ controls the between-class variability within each view.}	\label{tab: NoPrioS1}
	\end{footnotesize}	
\end{table}	

\begin{table}[http!]
	\begin{footnotesize}
		\begin{tabular}{lrrrrrrrr}
			\hline
			\hline
			Method&Error (\%)&  $\hat{\rho}$&	TPR-1&TPR-2 & FPR-1 &FPR-2  & F-1& F-2	\\
			\hline
			$(\rho_1=0.9, \rho_2=0.7, c=0.5)$\\
			SIDA (RS)&	2.16&	0.97&	83.75&	87.17&	0.19&	0.03&	85.40&	92.06\\
			SIDA (GS)&	2.25&	0.97&	84.38&	87.17&	0.20&	0.03&	85.64&	92.06\\
			sCCA&	3.61&	0.96&   	83.54&	88.04&	1.28&	6.23&	60.72&	52.73\\
			JACA&	2.08&	0.98&   	83.96&	87.61&	1.73&	1.91&	56.06&	54.12\\
			MGSDA (Stack)&	2.59&	0.83&	33.75&	31.09&	0.01&	0.02&	49.04&	45.97\\
			MGSDA (Ens) &	3.31&	0.93&	    46.25&	45.87&	0.09&	0.08&	59.25&	59.33\\
			\hline
			& &\\
			$(\rho_1=0.4, \rho_2=0.2, c=0.2)$\\
			SIDA (RS)&       22.8 	&0.40	&85.45	&82.95	&0.17	&0.18	&85.64	&85.60\\
			SIDA (GS)&   22.32	&0.40	&88.18	&87.27	&1.09	&1.03	&74.25	&74.37\\
			sCCA&       28.49	&0.49	&84.77	&85.68	&1.49	&1.28	&59.31	&60.43\\
			JACA&       20.77	&0.49	&91.14	&91.14	&1.01	&0.95	&72.83	&74.24\\
			MGSDA (Stack)&25.55	&0.34	&47.95	&45.91	&0.07	&0.11	&61.50	&58.65\\
			MGSDA (Ens)& 27.97	&0.39	&57.73	&57.73	&0.18	&0.39	&66.36	&62.03\\
			\hline
			& &\\
			$(\rho_1=0.15, \rho_2=0.05, c=0.12)$\\
			SIDA (RS)&       48.84	&0.03	&31.82	&44.29	&0.59	&2.29	&33.79	&33.28\\
			SIDA (GS)&   47.69	&0.03	&30.45	&44.76	&0.49	&1.72	&34.02	&34.40\\
			sCCA&       50.02	&0.03	&29.55	&42.14	&0.47	&1.31	&33.19	&36.54\\
			JACA &      40.42	&0.07	&63.64	&66.67	&1.03	&0.95	&56.00	&55.51\\
			MGSDA (Stack)&47.72 &0.03	&22.50	&25.48	&0.37	&0.41	&30.77	&33.18\\
			MGSDA (Ens)&49.77	&0.04	&26.36	&34.05	&0.74	&1.13	&32.70	&36.17\\
			\hline
			\hline
		\end{tabular}
		\caption{Scenario Two: We assume unequal covariances in each class. This violates the LDA assumption. RS; randomly select tuning parameters space to search. GS; search entire tuning parameters space.  MGSDA (Ens) applies sparse LDA method on separate views and peform classification on the pooled discriminant vectors. MGSDA (Stack) applies sparse LDA on stacked views. TPR-1; true positive rate for $\bX^1$. Similar for TPR-2. FPR; false positive rate for $\bX^2$. Similar for FPR-2; F-1 is F-measure for $\bX^1$. Similar for F-2. $\rho_1$ and $\rho_2$ controls the strength of association between $\bX^1$ and $\bX^2$. $c$ controls the between-class variability within each view. }	\label{tab: NoPrioS2}
	\end{footnotesize}	
\end{table}	

\begin{table}[http]
	\begin{footnotesize}
		\begin{tabular}{lrrrrrrrr}
			\hline
			\hline
			Method&Error (\%)&  $\hat{\rho}$&	TPR-1&TPR-2 & FPR-1 &FPR-2  & F-1& F-2	\\
			\hline
			$(\rho_1=0.9, \rho_2=0.7, c=0.25)$\\
			SIDA (RS)&	0.77&	0.91&	100.00&81.50&	0.13&	0.00&	96.14&	89.01\\
			SIDA (GS)&0.83&	0.90&   	99.50&	71.50&	0.13&	0.00&	95.84&	82.21\\
			sCCA&       1.08	&0.96	&97.75	&100.00	&0.06	&0.01	&96.41	&100.00\\
			JACA&	0.95&	0.96&   	100.00&	100.00&	0.34&	0.35&	89.14&	89.46\\
			MGSDA (Stack)&	1.78&	0.83&	17.25&	17.25&	0.02&	0.01&	39.57&	27.87\\
			MGSDA (Ens) &	1.36&	0.87&	    34.00&	25.00&	0.01&	0.02&	49.23&	37.18\\
			\hline
			& &\\
			$(\rho_1=0.4, \rho_2=0.2, c=0.2)$\\
			SIDA (RS)&       9.19 	&0.37	&58.00	&57.00	&1.39	&0.68	&56.46	&59.50\\
			SIDA (GS)&   9.28	&0.37	&60.75	&58.75	&1.55	&1.38	&51.75	&56.80\\
			sCCA&       9.81	&0.37	&56.75	&60.75	&0.00	&0.01	&71.35	&73.53\\
			JACA&       9.97	&0.40	&74.50	&79.00	&2.95	&2.56	&40.85	&47.27\\
			MGSDA (Stack)&10.75	&0.32	&18.00	&17.25	&0.13	&0.12	&27.00	&25.88\\
			MGSDA (Ens)& 12.95	&0.34	&21.00	&23.50	&0.10	&0.23	&31.34	&31.66\\
			\hline
			& &\\
			$(\rho_1=0.15, \rho_2=0.05, c=0.12)$\\
			SIDA (RS)&       23.83	&0.09	&50.00	&49.00	&1.87	&3.75	&47.25	&33.09\\
			SIDA (GS)&   23.38	&0.09	&51.00	&50.25	&2.63	&3.14	&41.21	&38.00\\
			sCCA&       27.69	&0.07	&37.50	&41.50	&5.30	&0.07	&49.75	&58.54\\
			JACA &      22.63	&0.10	&43.00	&42.50	&0.38	&0.16	&52.12	&54.36\\
			MGSDA (Stack)&24.77 &0.08	&13.00	&10.75	&0.12	&0.12	&21.15	&18.04\\
			MGSDA (Ens)&26.95	&0.08	&13.00	&10.75	&0.35	&0.14	&18.28	&17.43\\
			\hline
			\hline
		\end{tabular}
		\caption{Scenario Three: Binary class, equal covariance within class. RS; randomly select tuning parameters space to search. GS; search entire tuning parameters space.  MGSDA (Ens) applies sparse LDA method on separate views and peform classification on the pooled discriminant vectors. MGSDA (Stack) applies sparse LDA on stacked views. TPR-1; true positive rate for $\bX^1$. Similar for TPR-2. FPR; false positive rate for $\bX^2$. Similar for FPR-2; F-1 is F-measure for $\bX^1$. Similar for F-2. $\rho_1$ and $\rho_2$ controls the strength of association between $\bX^1$ and $\bX^2$. $c$ controls the between-class variability within each view. }	\label{tab: NoPrioS3}
	\end{footnotesize}	
\end{table}

\subsection{Example 2: simulation settings when prior information is available }\label{sec:simul2}
\begin{figure}[httb]\label{fig:sidanet}
	\begin{center}
		\begin{tabular}{l}
			\includegraphics[height = 3.5in]{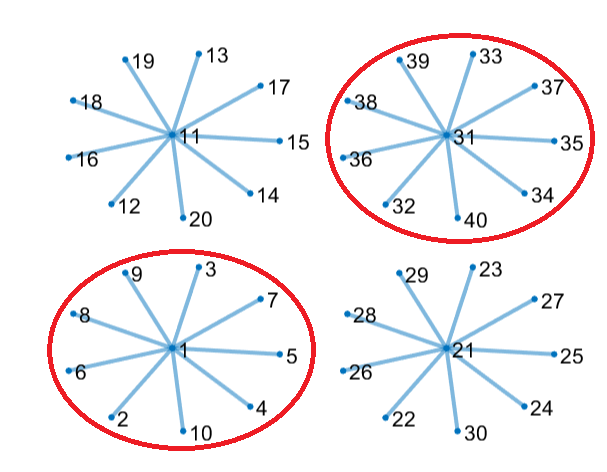}
		\end{tabular}
	\end{center}
	\caption{Simulation setup when network information is available. In scenario one, all four networks contribute to both separation and association. In the second scenario, two networks (circled) contribute to both separation and association.}
\end{figure}
In this setting, there are three views of data $\bX^d, d=1,2,3$, and each view is a concatenation of data from three classes. The true covariance matrix $\bSigma$ is defined as in Model 1 but with the following modifications. We include $\bSigma_{3}$ $\bSigma_{13}$, and $\bSigma_{23}$. $\tilde{\bSigma}^{1}$, $\tilde{\bSigma}^{2}$, and $\tilde{\bSigma}^{3}$ are each block diagonal with four blocks of size 10 representing four networks, between-block correlation 0, and  each block is a compound symmetric matrix with correlation 0.7. Each block has  a  9 $\times$ 9 compound symmetric submatrix with correlation 0.49 capturing the correlations between other variables within a network. The cross-covariance matrices $\bSigma_{12}$, $\bSigma_{13}$, and $\bSigma_{23}$  follow Model 1, but to make the effect sizes of the main variables larger, we multiply their corresponding values in $\bV^d,d=1,2,3$ by 10. We set $\bD=$diag$(0.9,0.7)$ when computing the cross-covariances.

We consider two scenarios in this example that differ by how the networks contribute to both separation and association.  In the first scenario, all four networks contribute to separation of classes within each view and association between the views. Thus, there are forty signal variables for each view, and $p_1-40, p_2-40$ and $p_3-40$ noise variables. In the second scenario, only two networks in the graph structure contribute to separation and association; hence there are twenty signal variables and $p_1-20, p_2-20$ and $p_3-20$ noise variables. Figure 3 is a pictorial representation for  the two scenarios. For each scenario, we set $n_k=40, k=1,2,3$ and generate the combined data $(\bX^1_k, \bX^2_k, \bX^3_k))$ from $MVN(\bmu_k. \bSigma)$. We set $c$ (refer to Model 1) to $0.2$ when generating the mean matrix $\bmu_k$.

\subsubsection{Competing Methods and Results}
We compare SIDANet with Fused Sparse LDA (FNSLDA) \citep{SAFOSADM2019}, a  classification-based method that incorporates prior information in sparse LDA. We apply FNSLDA on the stacked views [FNSLDA (Stack)] and use the classification algorithm proposed in the original paper. We also perform FNSLDA on separate views and perform classification on the combined discriminant vectors as described in Section 7 [FNSLDA (Ens)]. We perform FNSLDA using the Matlab code the authors provide, and use the default option for selecting the optimal tuning parameters. We did not find any comparable association-based method that incorporates prior information and is applicable when there are more than two views of data. We evaluate the methods using the same criteria listed in Section \ref{sec:evalcriteria}. 

Table \ref{tab:PriorS1} shows the performance of the proposed method compared to other methods. Compared to FNSLDA, SIDANet tends to have competitive TPR, lower FPR, higher F$_1$ scores, and competitive error rates and estimated correlations. These findings, together with the findings when there are no prior information, underscore the benefit of considering joint integrative and classification methods when the goal is to both correlate multiple views of data and perform classification simultaneously.

\begin{table}[httb]
	\begin{scriptsize}			\begin{tabular}{lrrrrrrrrrrr}
			\hline
			\hline
			Method&Error (\%)&  $\hat{\rho}$&	TPR-1&TPR-2 &TPR-3& FPR-1 &FPR-2 &FPR-3 & F-1& F-2& F-3	\\
			\hline
			\hline
			\textbf{Scenario One}&  \\
			\rb{SIDANet (RS)}	&1.57& 0.87&	99.88& 99.25& 98.00&	1.49& 4.12& 2.28&	87.79&67.88&80.24\\
			\rb{SIDANet (GS)}	&1.81& 0.87&	99.25& 98.88& 94.25&	1.92& 1.31& 0.92&	85.26&88.94&89.93\\
			
			FNSLDA (Ens)&	1.59&0.88	&100.00&100.00  &100.00 &7.25& 2.00 &2.83 &75.80&85.29&82.01\\
			FNSLDA (Stack)&	1.50&0.87	&100.00&100.00  & 100.00 &8.95&9.04	&8.81 &79.34& 78.67&	79.15\\
			\hline				
			\textbf{Scenario Two}&  \\
			\rb{SIDANet (RS)}	&3.69& 0.88&	99.50& 100.00& 91.75&	1.40& 2.31& 1.01&	78.85&65.16&	74.21\\
			\rb{SIDANet (GS)}	&3.78& 0.88&	100.00& 99.75& 86.50&	1.39& 0.95& 0.31&	79.18&86.05&	85.05\\
			FNSLDA (Ens)&	4.03&0.87	&100.00&100.00  &100.00 &7.01& 4.46 &12.55 &	52.43&52.91&44.25\\
			FNSLDA (Stack)&	3.73&0.85	&100.00&100.00  & 100.00 &16.63&16.46	&16.80 &38.52& 38.52&	38.49\\
			\hline
			\hline
		\end{tabular}
		\caption{Scenario One: all four networks contribute to separation of classes within each dataset, and association between the three views of data. Scenario Two: two networks contribute to both separation and association. FNSLDA (Ens) applies fused sparse LDA on separate views and perform classification on the combined discriminant vectors. FNSLDA (Stack) applies fused sparse LDA on stacked views. TPR-1; true positive rate for $\bX^1$. Similar for TPR-2 and TPR-3. FPR; false positive rate for $\bX^2$. Similar for FPR-2 and FPR-3; F-1 is F-measure for $\bX^1$. Similar for F-2 and F-3.  }	\label{tab:PriorS1}
	\end{scriptsize}	
\end{table}	

\section{Real data analysis}
We focus on analyzing the gene expression, metabolomics, and clinical data from the PHI study.  Our main goals are to i) identify genes and metabolomics features (mass-to-charge ratio [m/z]) that are associated and optimally separate subjects at high-vs low-risk for developing ASCVD, and ii) assess the added benefit of the identified variables in ASCVD risk prediction models that include some established risk factors (i.e., age and gender). 

\noindent\normalsize{\textit{\textbf{Data preprocessing and application of the proposed and competing methods}}}:\\
We use data for 142 patients for whom gene expression and metabolomics data are available and for whom there are clinical and demographic variables to compute ASCVD risk score. The ASCVD risk score for each subject is dichotomized  into high (ASCVD $>$ 5\%) and low (ASCVD $\le$ 5\%) risks based on guidelines from the American Heart Association.  The data consists of 87 females and 55 males; their ages range from 40 to 78 with mean age 53.8 years. The proportion of high and low risks  are respectively $80.3\%$ and $19.7\%$. The gene expressions data consist of $38,694$ probes, and the metabolomics data  consist of $\sim 6,000$ mass to ion (m/z) features. We preprocess and preselect genes as follows.  We remove genes with variance and entropy expression values that are respectively less than the 90th and 20th percentile, resulting in $1,658$ genes. We obtain the gene-gene interactions from the human protein reference database (HPRD) \citep{hprd2003}. The resulting network has 519 edges. For the metabolomics data, we removed m/z features with at least 50\% zeros, and features with coefficient of variation $\ge 50\%$; this resulted in 2,416 features for the analyses. Because of the skewed distributions of most metabolomic levels, we log2 transformed each feature. Both datasets are normalized to have mean 0 and variance 1 for each variable. We divide each view of data equally into training and testing sets. We select the optimal tuning parameters that maximize average classification accuracy from 5-fold cross validation on the training set. The selected tuning parameters are then applied to the testing set to estimate test classification accuracy. The process is repeated 20 times and we obtain average test error, variables selected, and RV coefficient.  \\
\noindent\normalsize{\textit{\textbf{Average misclassification rates, estimated correlations and variables selected}}}: Table \ref{tab:realresults3} shows the average results for the twenty resampled datasets. Of note, (+ covariates) refers to when the covariates age, gender, BMI, systolic blood pressure, low-density lipoprotein (LDL), and triglycerides  are added as a third dataset to SIDA or SIDANet; we assess the results with and without covariates. For SIDANet, we only incorporate prior network information from the gene expressions data (i.e., protein-protein interactions). sLDA (Ens) and sLDA (Stack) utilize the sparse linear discriminant method \citep{Irina:2015}. For sCCA, we utilize the sparse CCA method \citep{SafoBiomSELP} and obtain the first canonical vectors for the gene expression and metabolomics data. We combine the canonical vectors and use the pooled classification algorithm from Section 7 to classify. We also compute RV  coefficient using the canonical vectors and the training data. 

We observe that SIDA and SIDANet offer competitive results in terms of separation of the ASCVD risk groups. They also yield higher estimated correlations between the gene expressions and metabolomics data. SIDANet yields higher estimated correlation and competitive error rate when compared to SIDA, which suggests that incorporating prior network information may be advantageous. 
It seems that including  covariates in this example does not make the average classification accuracy and correlation any better. 
From this application, stacking the data results in better classification rate, but the estimated correlation is poor, which is not surprising since this approach ignores correlation that exists between the datasets. Among the methods compared, sLDA (Ens) and sLDA (Stack), which use the sparse LDA method in \cite{Irina:2015}, identify fewer number of genes and m/z features. This agrees with the results from the simulations where these methods had lower false and true positive rates. 
\begin{table}[htbp]
	\begin{center}
		\begin{small}
			\begin{tabular}{llrrrrr}
				\hline
				~	 &Error (\%)	& \# Genes &	\# m/z features&	Correlation \\
				\hline
				\rb{SIDA} &22.18     &193.80	&136.50 &	0.65\\
				\rb{SIDA (+ covariates)} &22.68     &60.75	&38.45 &	0.45\\
				\rb{SIDANet}&22.39&244.60 &165.40 & 0.70\\
				\rb{SIDANet (+ covariates)} &22.82     &63.65	&34.60 &	0.45\\
				sCCA &46.48 &139.75& 336.25& 0.43\\		
				JACA    & 25.49   	&637.20	&871.65&	0.52\\
				sLDA (Ens)& 30.28              &14.20	&11.60   &	0.23\\
				sLDA (Stack) &19.15 &4.25 & 6.20 &0.09\\		
				
				\hline
				\hline
			\end{tabular}
			\caption{SIDA (+covariates) uses RS and includes other covariates (see text) as a third dataset. 
				SIDANet uses prior network information from the gene expression data alone. sLDA (Ens) separately applies sparse LDA on the gene expression and metabolomics data and combines discriminant vectors when estimating classification errors. sLDA (Stack) applies sparse LDA on the stacked data. SIDA and SIDANet have competitive error rate and higher estimated correlations. It seems that including covariates does not make the average classification accuracy and correlation any better. 
				\label{tab:realresults3}}		
		\end{small}
	\end{center}
\end{table}

\noindent\normalsize{\textit{\textbf{Variable stability}}}:
To reduce false findings and improve variable stability, we use resampling techniques and consider two criteria to identify variables that potentially discriminate persons at high -vs low- risk for ASCVD. Specifically, out of the 20 resampled datasets, we chose variables that are selected at least 12 times ( $\ge$ 60\%), and which have average absolute coefficients within the top 1\%. From Table \ref{tab:realresults4}, SIDANet and JACA selected 14 genes, of which 8 overlap. Additionally, there are 9 overlapping genes and 6 m/z features for SIDA and SIDANet.   Meanwhile, JACA selects only 1 m/z feature while SIDANet and SIDA respectively select 6 and 9 m/z features. sLDA (Ens) and sLDA (Stack) did not identify any gene and m/z feature (refer to Tables 2 and 3 in supplemental materials).

\begin{table}[htbp]	
	\begin{center}
		\begin{small}
			\begin{tabular}{lcccc}
				\hline
				~	 	& \# Genes &	\#m/z features \\
				\hline
				\rb{SIDA}     &11	&9 \\	
				\rb{SIDANet}	&14 &6\\
				sCCA &1& 24 \\
				JACA  &14 &1\\	
				sLDA (Ens)  &0 &0\\
				sLDA (Stack)  &0 &0\\	
				\hline
				\hline
			\end{tabular}
			\caption{Genes and m/z feature selected at least 60\% (12 times out of 20 resampled datasets) and with average effect size within the top 1 \%. 
				\label{tab:realresults4}}				
		\end{small}
	\end{center}
\end{table}

\noindent\normalsize{\textit{\textbf{Genes or m/z features from SIDA and SIDANet plus established risk factors predict ASCVD better}}}: Our aim here is to assess whether including the genes or m/z features identified by our methods is any better than a model with only age and gender. Given the sample size of 71 in each of the 20 testing resampled datasets, we can only include a few variables to increase power of detecting  differences in low vs high-risk ASCVD. We  include the demographic variables age and gender in model one (M1). In model two we further include a risk score calculated with the genes or m/z features identified by the methods using the testing datasets. Specifically, we run a logistic regression model on the training data to obtain effect sizes (logarithm of the odds ratio of the probability that ASCVD risk group is high) for each gene or m/z feature. The genetic risk score (GRS) or metabolomic risk score (MRS) are each obtained  as a sum of the genes or m/z features in the testing data set, weighted by the effect sizes. In Model 3 (M3), we include both GRS and MRS. We summarize the area under the curves (AUCs) from the receiver operating characteristic in Table \ref{tab:realresults6}. We observe that including genes and/or m/z features identified by our methods to a model with age and gender results in better discrimination of the ASCVD risk groups compared to association or classification-based methods, and when compared to a model with only age and gender. By integrating gene expression and m/z features and simultaneously discriminating ASCVD risk group, we have identified biomarkers that potentially may be used to predict ASCVD risk, in addition to a few established ASCVD risk factors. 

\begin{table}[htbp]
	\begin{center}
		\begin{small}
			\begin{tabular}{llrrrrr}
				\hline
				~	 &minimum	& mean  &	median &	maximum \\
				\hline
				M1 &0.71 &0.80 & 0.81&0.89\\
				& &\\
				\hline
				M2: M1 + GRS & &\\
				\rb{SIDA} &0.81     &0.89	&0.90 &	0.95\\
				\rb{SIDANet} &0.82	&0.91	&0.91&	0.96\\				JACA    & 0.83   	&0.93	&0.94&	0.99\\
				sCCA &0.71 &0.81& 0.82& 0.90\\	
				& &\\			
				\hline
				M3: M1 + MRS & &\\
				\rb{SIDA} &0.80     &0.87	&0.87 &	0.97\\
				\rb{SIDANet}&0.79&	0.86&	0.86&	0.97\\
				JACA    & 0.78   	&0.85	&0.85&	0.91\\
				sCCA &0.72 &0.81& 0.82& 0.89\\
				& &\\
				\hline
				M4: M1 + GRS + MRS & &\\
				\rb{SIDA} &0.87     &0.93	&0.93 &	0.99\\
				\rb{SIDANet}&0.85&	0.93&	0.93&	0.97\\
				JACA    & 0.84   	&0.95	&0.96&	0.99\\
				sCCA &0.72 &0.82& 0.82& 0.90\\				
				\hline
				\hline
			\end{tabular}
			\caption{ Comparison of AUCs using genes and m/z features identified: Model 1 (M1): Age + gender;  Model 2 (M2): Age + gender + gene risk score (GRS); Model 3 (M3): Age + gender + metabolomic risk score (MRS).  Model 4 (M4): Age + gender + gene risk score (GRS) + metabolomic risk score (MRS). The genes and m/z features identified by the methods on the training datasets are used to calculate GRS and MRS. Summary statistics are over 20 AUCs.  \label{tab:realresults6}}	
			
		\end{small}
	\end{center}
\end{table}


\section{Conclusion}
We have proposed two methods for joint integrative analysis and classification studies to add to the limited literature in this area. One of the methods proposed here is  both data- and knowledge-driven and useful when prior biological information about variable-variable interactions is available. The numerical experiments and the data analyses described in this paper underscore the benefit of joint integrative and classification analysis methods when the goal is to correlate multiple views of data and to perform classification simultaneously. The encouraging findings from the real data analysis motivate further applications. We acknowledge some limitations in our methods. The methods we propose are only applicable to complete data and do not allow for missing values. A future project could extend the current methods to the scenario where data are missing using multiple imputation methods. We assume equal contributions of separation and association to the overall optimization problem.  It would be interesting to consider the performance of the proposed methods when this parameter is allowed to vary, or is chosen in a data-adaptive way. 
\section*{Acknowledgements}
We are grateful to the Emory Predictive Health Institute for providing
us with the gene expression, metabolomics, and clinical data.  This research is partly supported by NIH grants 1KL2TR002492\-01 and  T32HL129956. The content is solely the responsibility of the authors and does not necessarily represent the official views of the NIH.
\section*{Supplemental Material}
In the online Supplemental Materials, we provide proof for Theorem 1. We provide a detailed comparison of \textit{random} and \textit{grid} search in terms of error rates, estimated correlations, variables selected, and computational times.
Matlab and R codes for implementing the methods along with README files may be found on the corresponding author's website. 

\newpage


\newpage
\section{Supplementary Material}
\subsection{Proof of Theorem 1}
The Lagrangian 
\begin{eqnarray*}
	L(\bA, \bB,\lambda_1,\lambda_2) 
	&=& \rho\text{tr}(\bA^{{\smt}}\bS_{b}^{1} \bA+  \bB^{{\smt}}\bS_{b}^{2} \bB) + (1-\rho) \text{tr}(\bA^{{\smt}}\bS_{12}\bB\bB^{{\smt}}\bS_{12}^{\smt}\ \bA) \nonumber\\
	&-& \lambda_1(\text{tr}(\bA^{{\smt}}\bS_{w}^{1} \bA)-(K-1)) - \lambda_2(\text{tr}(\bB^{{\smt}}\bS_{w} \bB)- (K-1))\nonumber\\
\end{eqnarray*}

Let $\bOmega^1= \bS_{12}\bB\bB^{{\smt}}\bS_{12}^{\smt}$ and $\bOmega^2=\bS_{12}^{\smt}\bA\bA^{{\smt}}\bS_{12}$.\\
\noindent The first order stationary solutions for $\bA$ and $\bB$ are\\
\begin{eqnarray}
\frac{\partial{L(\bA, \bB,\lambda_1,\lambda_2)}}{\partial{\bA}} = \rho(\bS_{b}^{1} + \bS_{b}^{1^{\smt}})\bA + (1-\rho)(\bOmega^1 + \bOmega^{1^{\smt}})\bA - \lambda_1(\bS_{w}^{1}  + \bS_{w}^{1^{\smt}})\bA = \mathbf{0}\nonumber\\
\frac{\partial{L(\bA, \bB,\lambda_1,\lambda_2)}}{\partial{\bB}} = \rho(\bS_{b}^{2} + \bS_{b}^{2^{\smt}})\bB + (1-\rho)(\bOmega^2 + \bOmega^{2^{\smt}})\bB - \lambda_1(\bS_{w}^{2}  + \bS_{w}^{2^{\smt}})\bB = \mathbf{0}\nonumber\
\end{eqnarray}
Rearranging, we obtain the eigensystems for $\bA$ and $\bB$ respectively as\\
\begin{eqnarray}
\left(\rho(\bS_{b}^{1} + \bS_{b}^{1^{\smt}}) + (1-\rho)(\bOmega^1 + \bOmega^{1^{\smt}})\right)\bA &=& \lambda_1(\bS_{w}^{1}  + \bS_{w}^{1^{\smt}})\bA \\
\left(\rho(\bS_{b}^{2} + \bS_{b}^{2^{\smt}}) + (1-\rho)(\bOmega^2 + \bOmega^{2^{\smt}})\right)\bB &=& \lambda_2(\bS_{w}^{2}  + \bS_{w}^{2^{\smt}})\bB 
\end{eqnarray}
For $\bB$ fixed in $\bOmega^1$, equation (1) can be solved for the nonzero eigenvalues of  $(\bS_{w}^{1}  + \bS_{w}^{1^{\smt}})^{-1}(\rho(\bS_{b}^{1} + \bS_{b}^{1^{\smt}}) + (1-\rho)(\bOmega^1 + \bOmega^{1^{\smt}}))$. Denote the corresponding eigenvectors as $\widetilde{\bA}=[\tilde{\balpha}_1,\ldots,\tilde{\balpha}_r]$. Similarly, with $\bA$ fixed in $\bOmega^2$, we can solve for the nonzero eigenvalues in equation (2) from $(\bS_{w}^{2}  + \bS_{w}^{2^{\smt}})^{-1}(\rho(\bS_{b}^{2} + \bS_{b}^{2^{\smt}}) + (1-\rho)(\bOmega^2 + \bOmega^{2^{\smt}}))$. Let $\widetilde{\bB}=[\tilde{\bbeta}_1,\ldots,\tilde{\bbeta}_r]$. We iterate over $\bA$ and $\bB$ in equations (1) and (2) until convergence (both $\|\widetilde{\bA}_{new} -\widetilde{\bA}_{old}\|_{F}< \epsilon$ and $\|\widetilde{\bB}_{new} -\widetilde{\bB}_{old}\|_{F}< \epsilon$).
At which point we set $\widehat{\bA}=\widetilde{\bA}$ and $\widehat{\bB}=\widetilde{\bB}$. 

\subsection{Time Comparisons}
We compare the run times of  \textit{random } and \textit{grid search}.   We consider a $K=3$ class and $D=2$ views problem  and simulate data according  to Scenario One in the main text when no prior information exists. In \textit{grid search}, we choose tuning parameters over a $8 \times 8$ grid (or 64 grid points). \textit{Random search} randomly selects $15\%$ of the grid points to optimize. We compare run times for $N < p$ and $N>p$, and when the cross validation task for choosing optimal tuning parameters is executed in parallel (using 4 workers) or not. All comparisons are carried out with the Matlab codes for SIDA on an Intel (R) Core (TM) i7-7700 3.60 GHz processor. Table 1 gives timings in minutes averaged over three runs. We see that \textit{random search} is considerably faster than \textit{grid search}. SIDA with \textit{random search}, with or without parallelization is faster than JACA especially when $N<p$. 
\begin{table}[htbp!]
	\begin{center}
		\begin{small}
			\begin{tabular}{llllll}
				\hline
				~	 &SIDA (RS, P)	& SIDA (GS, P)&	SIDA (RS, NP)& SIDA (GS, NP) &JACA  \\
				\hline
				($N$, $p/q$)&\\
				(240, 200/200) &1.49     &6.80	&8.43 &	39.79 &1.31\\
				(240, 2000/2000) &3.39     &13.32	&12.90 &	61.51&22.31\\
				
				(1000, 200/200) &1.36     &6.52	&10.24 &	35.00 &3.22\\
				(1000, 2000/2000) &5.61     &26.35	&12.81 &	66.31 &69.53\\
				\hline
				\hline
			\end{tabular}
			\caption{Timings (in minutes). Average time for five fold cross-validation.  RS and GS denote \textit{random} and \textit{grid} search respectively. P is parallel computing (4 workers), and NP is no parallel computing.  $N$ is the sample size, and $p/q$ are the dimensions for the two views of data. 
				\label{tab:srealresults3}}		
		\end{small}
	\end{center}
\end{table}

\subsection{Real Data Analysis}
\subsubsection{Genes and m/z features selected by methods}
Tables 2 and 3 give the genes and m/z features selected by the proposed and competing methods at at least 60\% (12 times out of 20 resampled datasets) and with average effect size within the top 1 \%.  SIDANet and JACA selected 14 genes, of which 8 overlap. Additionally, there are 9 overlapping genes and 6 m/z features for SIDA and SIDANet.   Meanwhile, JACA selects only 1 m/z feature while SIDANet and SIDA respectively select 6 and 9 m/z features. sLDA (Ens) and sLDA (Stack) did not identify any gene and m/z feature.
\begin{table}
	\begin{center}
		\begin{scriptsize}
			\begin{tabular}{ll}
				\hline\\
				Method &Genes selected\\
				\hline
				SIDA&	DEFB127	ERV3	GLYAT		H3F3A	HIST1H2BG HIST1H4H	MAGEB4\\
				&RASEF	SCGB1C1	SCUBE1	TENC1	\\		
				SIDANet&	BZRAP1	CIRBP	CLEC1B	CYP17A1	ERV3	H3F3A	HIST1H2BG\\
				&HIST1H4H	HMBOX1	MAGEB4	RASEF	SCGB1C1	SCUBE1	TENC1\\	
				sCCA&	PSMA3	\\													
				JACA&	ABHD3	CIRBP	CYP17A1		DARC	ERV3	GLYAT	H3F3A\\
				&HIST1H2BG	HIST1H4H	MAGEB4	NEURL2	PTGS2	RASEF	SCGB1C1\\
				sLDA(Ens)& -				\\											
				sLDA(Stack)	& -		\\												
				\hline
				\hline
			\end{tabular}
			\caption{Genes feature selected at least 60\% (12 times out of 20 resampled datasets) and with average effect size within the top 1 \%. There are nine overlapping genes between SIDA and SIDANet.
				\label{tab:srealresults5a}}				
		\end{scriptsize}
	\end{center}
	\begin{center}
		\begin{scriptsize}
			\begin{tabular}{ll}
				\hline\\
				Method &m/z features (retention times) selected\\
				\hline
				SIDA& 	168.9045(	73.1430)	212.9862	(373.9647)	216.9397(	134.2085)	228.8127	(98.0079)	250.1187	(30.9802)	\\
				&342.3191	(37.0602)	542.3191 (572.5522)	754.4435	(42.6461)	756.7378	(64.1087)	\\								
				SIDANet&	168.9045	(73.1430)	216.9397	(134.2085)	250.1187	(30.9802)	542.3191	(572.5522)	754.4435	(42.6461)	\\	
				&756.7378	(64.1087)	\\	
				sCCA&	89.0796	(37.7417)	136.0216	(42.0915)	153.1274	(552.8494)		201.2042	(35.9136) 226.8615	(64.5573)	\\
				&	234.2039	(439.2704)	238.2159	(587.6828)	249.1846	(27.0957) 284.2946	(596.0450)	295.2263	(593.7375) 	\\
				&404.1029	(517.3633)	461.3614	(36.8199)	509.8287	(42.4354) 553.3890	(51.6947) 561.3572 (51.5734) \\
				&		694.4398	(51.3845)	709.4125	(42.9693)	738.4680	(51.1611)		739.4723	(51.2275) 753.4380  (41.6097)\\
				&	797.4647	(42.4703)	826.5196	(49.9539)	841.4923	(42.3359
				869.5445	(50.2823)		\\			
				JACA&	102.0666	(140.1219)	\\	
				sLDA(Ens)& -				\\											
				sLDA(Stack)	& -		\\	
			\end{tabular}
			\caption{m/z features (retention times) selected at least 60\% (12 times out of 20 resampled datasets) and with average effect size within the top 1 \%. There are six overlapping features between SIDA and SIDANet. 
				\label{tab:srealresults5b}}				
		\end{scriptsize}
	\end{center}
\end{table}

\
\subsubsection{Comparison of Genes and m/z features selected by SIDA and SIDANet for both random and grid search}
We compare genes and m/z features identified by SIDA and SIDANet using both random search and grid search for tuning parameter optimizations.  Table 4 gives the average error rate on the testing data, average estimated correlation on the training data, and average number of genes and m/z features. Averages are over 20 resampled datasets. SIDA with \textit{random search} and \textit{grid search} yield similar error rates, and estimated correlation. This is also true for SIDANet.  In terms of variable selected using the criteria discussed in the main text, eight genes and five m/z features overlap between SIDA with \textit{random} and \textit{grid} search (Table 6). Comparing SIDANet (RS) with SIDANet (GS), the 11 genes identified by SIDANet (GS) is a subset of the genes identified by SIDANet (RS) [ Table 6]. This is also true for the m/z features identified by SIDANet (RS) and SIDANet (GS) [Table 7]. Table 8 compares the AUC's for the three models under consideration. The results are simlar for both RS and GS. These findings suggest that we can choose optimal tuning parameters at a lower computational cost (see Table 1) by randomly selecting grid points from the entire tuning parameter hyperspace and searching over those grid values (instead of searching over the entire grid space) and still achieve competitive performace. In our algorithm, the default method to obtain optimal tuning parameter is \textit{random search}. However, we make it as an option for the interested user to choose tuning parameters using \textit{grid search}.
\begin{table}[htbp]
	\begin{center}
		\begin{small}
			\begin{tabular}{lllll}
				\hline
				~	 &Error (\%)	& \# Genes &	\# m/z features&	Correlation \\
				\hline
				\rb{SIDA (RS)} &22.18     &193.80	&136.50 &	0.65\\
				\rb{SIDA (GS)} &22.04     &179.35	&134.50 &	0.60\\
				\rb{SIDANet (RS)}&22.39&244.60 &165.40 & 0.70\\
				\rb{SIDANet (GS)} &22.46     &217.25	&152.55 &	0.59\\
				\hline
				\hline
			\end{tabular}
			\caption{SIDANet uses prior network information from the gene expression data alone.   
				\label{tab:ssrealresults3}}		
		\end{small}
	\end{center}
	\begin{center}
		\begin{small}
			\begin{tabular}{lcccc}
				\hline
				~	 	& \# Genes &	\# m/z features \\
				\hline
				{SIDA (RS)}     &11	&9 \\
				{SIDA (GS)}     &10	&5 \\
				{SIDANet (RS)}	&14 &6\\
				{SIDANet (GS)}	&11 &4\\				
				\hline
				\hline
			\end{tabular}
			\caption{Genes and m/z feature selected at least 60\% (12 times out of 20 resampled datasets) and with average effect size within the top 1 \%.  Eight genes and five  m/z features overlap between SIDA (RS) and SIDA (GS).
				\label{tab:srealresults4}}				
		\end{small}
	\end{center}
\end{table}

\begin{table}[htbp]	
	\begin{center}
		\begin{scriptsize}
			\begin{tabular}{ll}
				\hline\\
				Method &Genes selected\\
				\hline
				SIDA (RS)&	DEFB127	ERV3	GLYAT		H3F3A	HIST1H2BG HIST1H4H	MAGEB4\\
				&RASEF	SCGB1C1	SCUBE1	TENC1	\\		
				SIDA (GS) &BZRAP1 CBS CIRBP EMP2 HIST1H4H HMBOX1 MAGEB4 \\
				& RASEF SCUBEI TENCI  \\
				SIDANet (RS)&	BZRAP1	CIRBP	CLEC1B	CYP17A1	ERV3	H3F3A	HIST1H2BG\\
				&HIST1H4H	HMBOX1	MAGEB4	RASEF	SCGB1C1	SCUBE1	TENC1\\	
				SIDANet (GS)& BZRAP1	CIRBP	CLEC1B	CYP17A1	ERV3 HIST1H4H	HMBOX1		\\
				&	MAGEB4	RASEF		SCUBE1	TENC1\\	
				\hline
				\hline
			\end{tabular}
			\caption{Genes feature selected at least 60\% (12 times out of 20 resampled datasets) and with average effect size within the top 1 \%. Eight genes and 5 m/z features overlap between SIDA (RS) and SIDA (GS).
				\label{tab:realresults5a}}				
		\end{scriptsize}
	\end{center}
	\begin{center}
		\begin{scriptsize}
			\begin{tabular}{ll}
				\hline\\
				Method &m/z features (retention times) selected\\
				\hline
				SIDA (RS)& 	168.9045(	73.1430)	212.9862	(373.9647)	216.9397(	134.2085)	228.8127	(98.0079)	250.1187	(30.9802)	\\
				&342.3191	(37.0602)	542.3191 (572.5522)	754.4435	(42.6461)	756.7378	(64.1087)	\\	
				SIDA (GS)& 	168.9045(	73.1430)	216.9397(	134.2085)	250.1187	(30.9802)	\\
				&	754.4435	(42.6461)	756.7378	(64.1087)	\\
				SIDANet (RS) &	168.9045	(73.1430)	216.9397	(134.2085)	250.1187	(30.9802)	542.3191	(572.5522)	754.4435	(42.6461)	\\	
				&756.7378	(64.1087)	\\
				SIDANet (GS) &168.9045	(73.1430)	216.9397	(134.2085)	250.1187	(30.9802)		754.4435	(42.6461)	\\	
			\end{tabular}
			\caption{m/z features (retention times) selected at least 60\% (12 times out of 20 resampled datasets) and with average effect size within the top 1 \%. There are five overlapping features between SIDA (RS) and SIDA (GS). 
				\label{tab:realresults5b}}				
		\end{scriptsize}
	\end{center}
\end{table}

\begin{table}[htbp]
	\begin{center}
		\begin{small}
			\begin{tabular}{llrrrrr}
				\hline
				~	 &minimum	& mean  &	median &	maximum \\
				\hline
				M1 &0.71 &0.80 & 0.81&0.89\\
				& &\\
				\hline
				M2: M1 + GRS & &\\
				\rb{SIDA (RS) } &0.81     &0.89	&0.90 &	0.95\\
				\rb{SIDA (GS) } &0.82     &0.92	&0.92 &	0.97\\
				\rb{SIDANet (RS)} &0.82	&0.91	&0.91&	0.96\\	
				\rb{SIDANet (GS)} &0.82	&0.90	&0.91&	0.95\\	
				& &\\			
				\hline
				M3: M1 + MRS & &\\
				\rb{SIDA (RS)} &0.80     &0.87	&0.87 &	0.97\\
				\rb{SIDA (GS) } &0.79     &0.86	&0.85 &	0.94\\
				\rb{SIDANet (RS)}&0.79&	0.86&	0.86&	0.97\\
				\rb{SIDANet (GS)}&0.79&	0.86&	0.85&	0.93\\
				\hline
				& & \\
				M4: M1 + GRS + MRS & &\\
				\rb{SIDA (RS) } &0.87     &0.93	&0.93 &	0.99\\
				\rb{SIDA (GS) }&0.88&	0.93&	0.93&	0.98\\
				\rb{SIDANet (RS)}&0.85&	0.93&	0.93&	0.97\\
				\rb{SIDANet (GS)}&0.87&	0.92&	0.92&	0.97\\
				
				\hline
			\end{tabular}
			\caption{ Comparison of AUCs using genes and m/z features identified: Model 1 (M1): Age + gender;  Model 2 (M2): Age + gender + gene risk score (GRS); Model 3 (M3): Age + gender+ metabolomic risk score (MRS). Model 4 (M4): age + gender + metabolomic risk score + gene risk score. The genes and m/z features identified by the methods on the training datasets are used to calculate GRS and MRS. Summary statistics are over 20 AUCs.  \label{tab:srealresults6}}	
			
		\end{small}
	\end{center}
\end{table}
\clearpage
\end{document}